\documentclass[aps,prr,reprint,longbibliography,twocolumn,a4paper,superscriptaddress]{revtex4-2}
\usepackage{graphicx,caption,subcaption,amsmath,amsfonts,amssymb,multirow,extarrows,bm,acronym,float}
\usepackage[colorlinks,linkcolor=blue,citecolor=blue,urlcolor=blue ]{hyperref}
\floatstyle{plaintop}
\restylefloat{table}
\usepackage{CJKutf8}
\newcommand{\fs}{\mathcal{F}\text{-statistic}}

\newcommand{\Msun}{\,{\rm M}_\odot}

\newcommand{\DLUT}{School of Physics, Dalian University of Technology, Dalian 116024, People's Republic of China}

\def\aj{Astronomical Journal}                 
\def\apj{Astrophysical Journal}                

\def\prd{Phys.~Rev.~D}       
\def\prl{Phys.~Rev.~Lett}    
\def\aap{A\&A}                

\begin{document}

\title{First Overtone and Higher-Order Modes in the Ringdown Signal of GW231028}

\author{Hai-Tian Wang (\begin{CJK}{UTF8}{gbsn}王海天\end{CJK})}
\email{wanght9@dlut.edu.cn}
\affiliation{\DLUT}

\date{\today}

\begin{abstract}
The properties of a remnant black hole can be probed by analyzing the gravitational waves emitted during its ringdown phase. This signal provides a direct test of general relativity in the strong-field regime. In this study, we apply both a time-domain $\fs$ framework and full Bayesian time-domain sampling to the ringdown of GW231028\_153006. We report decisive evidence for multimode content, specifically identifying both the first overtone ($\ell|m|n=221$) and the higher order ($\ell|m|n=210$). The detection of the $221$ mode is statistically significant, achieving a Bayes factor of $\sim 189.2$ and an amplitude exclusion from zero at $>7\sigma$ credibility for an analysis beginning at $10\,M$ postpeak. These findings are rigorously validated against the numerical relativity injection SXS:BBH:1282, which confirms that such multimode features are physically expected for a remnant with the inferred high spin and mass ratio. The inclusion of the overtone mode allows for precise constraints on the remnant's properties, yielding a redshifted final mass of $246.2^{+22.3}_{-22.4}\,\Msun$ and a final spin of $0.81_{-0.10}^{+0.07}$ (at $90\%$ credibility), consistent with full inspiral-merger-ringdown predictions. A test of the no-hair theorem, enabled by this robust multimode detection, shows consistency with general relativity.
\end{abstract}

\maketitle

\acrodef{GW}{gravitational wave}
\acrodef{LIGO}{Laser Interferometer Gravitational-Wave Observatory}
\acrodef{LVKC}{LIGO-Virgo-KAGRA Collaboration}
\acrodef{LVC}{LIGO-Virgo Collaboration}
\acrodef{NR}{numerical relativity}
\acrodef{FD}{frequency-domain}
\acrodef{TD}{time-domain}
\acrodef{BH}{black hole}
\acrodef{TTD}{traditional time-domain}
\acrodef{BBH}{binary black hole}
\acrodef{GR}{general relativity}
\acrodef{PN}{post-Newtonian}
\acrodef{SNR}{signal-to-noise ratio}
\acrodef{PSD}{power spectral density}
\acrodef{PDF}{probability density function}
\acrodef{ACF}{auto-covariance function}
\acrodef{IMR}{inspiral-merger-ringdown}
\acrodef{QNM}{quasinormal mode}

\section{Introduction}\label{sec:intro}
The final phase of a \ac{BBH} coalescence, known as the ringdown, provides a direct probe of spacetime in the extreme gravity limit \citep{Schw_PRD_Vishveshwara1970, GW_APJL_Press1971, QNM_APJ_Teukolsky1973}. During this stage, the newly formed remnant \ac{BH} settles to equilibrium by emitting \acp{GW} that can be described as a superposition of damped sinusoids, or \acp{QNM} \citep{Berti:2009kk}. Each \ac{QNM} is characterized by angular indices $(\ell m)$ and an overtone index $n$, with the fundamental mode, $\ell|m|n=220$, typically being the most prominent.

A central tenet of \ac{GR}, the no-hair theorem, posits that the entire \ac{QNM} spectrum of a Kerr \ac{BH} is uniquely determined by just two parameters: its mass and spin \citep{Hawking:1971vc, PhysRevLett.34.905, kerr1963gw}. Verifying this theorem requires the confident detection of at least two distinct modes from a single event to perform a consistency check on the inferred remnant properties \citep{Cardoso:2019rvt,Berti:2025hly}. However, detecting modes beyond the fundamental mode-such as overtones $(n\geq 1)$ or sub-dominant modes $(\ell |m|\neq 22)$ \citep{Sperhake:2007gu}—is notoriously difficult due to the typically low \ac{SNR} of the ringdown signal.

Furthermore, a key challenge in ringdown analysis is selecting an appropriate start time. Analyses initiated too close to the merger peak risk being contaminated by non-linear dynamics and transient effects not modeled by linear perturbation theory. Recent numerical relativity studies suggest that analyses should begin at least $\sim8\,M$ after the peak to ensure the system is in the linear regime, especially when searching for higher overtones \citep{Baibhav:2023clw,Nee:2023osy,Zhu:2023mzv,Clarke:2024lwi,Giesler:2024hcr}, where $M$ is the redshifted remnant mass.

Within this constraint, definitive detections of a second \ac{QNM} have remained elusive until very recently. 
The prospects of such a detection are fundamentally determined by the \ac{SNR} contained specifically within the ringdown phase. While exceptionally loud inspiral-dominated signals, such as GW250114 (with a total \ac{SNR} $\gtrsim 80$), contain sufficient power to reveal higher modes \citep{KAGRA:2025oiz,LIGOScientific:2025obp}, typical events with moderate total \ac{SNR}s are often challenging if the signal power is distributed primarily in the inspiral. 

For instance, inspiral-dominated events with total \ac{SNR}s around $\sim 30$ to $40$ may still lack the necessary ringdown loudness for overtone detection \citep{Tang:2025jyj,LIGOScientific:2025cmm}.
The search for the first overtone mode in the ringdown of GW150914 has been a subject of active debate \citep{CalderonBustillo:2020rmh,2021PhRvD.103l2002A,2022PhRvL.129k1102C,Finch:2022ynt,Correia:2023bfn,Carullo:2023gtf,Wang:2023mst, Wang:2024yhb,Chandra:2025ipu}. While \citet{2022PhRvL.129k1102C} initially reported evidence for the first overtone, subsequent analyses have challenged this finding \citep{Correia:2023bfn}. Recent investigations suggest that this discrepancy may stem from data conditioning choices, particularly the downsampling process \citep{Wang:2023mst,Siegel:2024jqd,Berti:2025hly}. It has been shown that spectral leakage near the Nyquist frequency in downsampled data can artificially affect the noise power spectrum estimation, potentially inflating the significance of the overtone. Analyses employing robust downsampling techniques or higher sampling rates (e.g., $16$ kHz) typically yield results consistent with a non-detection of the overtone in GW150914 \citep{Wang:2023mst,Wang:2024yhb}.

In contrast, merger-ringdown dominated events, such as GW190521 \citep{2020PhRvL.125j1102A} and GW231123 \citep{LIGOScientific:2025rsn}, concentrate a larger fraction of their energy into the post-merger phase. This characteristic makes them efficient targets for spectroscopy even at moderate total \ac{SNR}s.
For GW190521, an initial claim of a sub-dominant mode \citep{Capano:2021etf} was found to weaken considerably when the analysis was restricted to the later, linear regime \citep{Siegel:2023lxl,Gennari:2023gmx}, a result attributable to the event's limited ringdown \ac{SNR} since the total signal \ac{SNR} is only approximately $14.7$ \citep{2020PhRvL.125j1102A}.
More recently, the analysis of GW231123, which also had a high remnant mass and ringdown \ac{SNR}, yielded the first decisive evidence for a higher-order mode starting well within this linear regime \citep{Wang:2025rvn}.
For GW231123, the \ac{SNR} of the full signal is approximately $20.7$, and the redshifted remnant mass reaches as high as $298.0\,\Msun$.

The event GW231028\_153006 (hereafter GW231028) was observed by the LIGO Hanford and Livingston detectors \citep{LIGOScientific:2025slb}. 
Based on the parameter estimation results from a full \ac{IMR} analysis performed by the LIGO-Virgo-KAGRA (LVK) Collaboration using the SEOBNRv5PHM waveform model \citep{Ramos-Buades:2023ehm}, the signal has a network optimal \ac{SNR} of \( 22.8^{+1.7}_{-1.6} \) (here, and throughout this work, we present the median value and uncertainties based on the $90\%$ credible interval).
The source properties are characterized by a mass ratio of $q={0.4^{+0.5}_{-0.2}}$ and significant spin, with an effective inspiral spin parameter of $\chi_{\mathrm{eff}}={0.46^{+0.13}_{-0.15}}$ and an effective precession spin of $\chi_{p}={0.55^{+0.25}_{-0.26}}$. The remnant black hole formed from the merger has a redshifted final mass of $247.3^{+17.0}_{-16.6}\,M_\odot$ and a final dimensionless spin of $0.85^{+0.04}_{-0.06}$, making it an excellent candidate for \ac{BH} spectroscopy.

In this work, we analyze the ringdown signal of GW231028 using the $\fs$ method. This framework, originally developed for other \ac{GW} sources like continuous waves \citep{Jaranowski:1998qm,Cutler:2005hc,Dreissigacker:2018afk,Sieniawska:2019qnx} and extreme mass-ratio inspirals \citep{Wang:2012xh}, has been proven to be a robust and efficient tool for ringdown analysis \citep{Wang:2024jlz,Wang:2024yhb,Dong:2025igh,Wang:2025rvn}. It enhances search efficiency by analytically maximizing the likelihood over the linear \ac{QNM} parameters (amplitudes and phases), thereby reducing the dimensionality of the parameter space.

Here, we report decisive evidence for multimode features in GW231028, supporting the presence of both the first overtone ($221$) and the sub-dominant mode ($210$). Specifically, the $\fs$ analysis yields a Bayes factor of $\mathcal{B}^{220+221}_{220}\sim 189.2$ at $\Delta t=10\,M$ postpeak, and $\mathcal{B}^{220+210}_{220}\sim 7.3\times 10^5$ at $\Delta t=8\,M$. The robustness of these findings is corroborated by two independent tests: a parallel analysis using the \ac{TTD} method, and an injection study using the \ac{NR} waveform SXS:BBH:1282 \citep{Boyle:2019kee,Scheel:2025jct} which mimics the properties of GW231028. In both checks, we recover consistent conclusions. The specific start times for the reported mode combinations were selected because the inferred remnant mass and spin at these times exhibit better consistency with the predictions from full \ac{IMR} waveform models such as NRSur7dq4 \citep{Varma:2019csw} and SEOBNRv5PHM \citep{Ramos-Buades:2023ehm}. Furthermore, subsequent tests of the no-hair theorem based on both detections using the $\fs$ method reveal no deviation from the predictions of \ac{GR}.


\section{Method}\label{sec:method}
Our analysis models the ringdown signal as a superposition of \acp{QNM}, which are the characteristic oscillations of the remnant \ac{BH}. Each \ac{QNM}, indexed by angular numbers $(\ell,m)$ and an overtone number $n=0,1,2\ldots$, is a damped sinusoid. Assuming that the remnant is a Kerr \ac{BH}, the oscillation frequency $f_{\ell m n}$\footnote{By default, the oscillation frequency $f_{\ell m n}$ and final mass $M_f$ are all defined in the detector frame.} and damping time $\tau_{\ell m n}$ of every mode are uniquely determined by the remnant's redshifted mass $M_f$ and dimensionless spin $\chi_f$ \citep{Leaver:1985ax, Berti:2005ys, Berti:2009kk}. The other two parameters describing each mode are its amplitude $A_{\ell m n}$ and phase $\phi_{\ell m n}$.

The complete time-domain waveform is constructed from the contributions of all excited modes through its polarization components, $h_+(t)$ and $h_\times(t)$:
\begin{equation}
\begin{aligned}
&h_+(t) + i h_{\times}(t) \\
&= \sum_{\ell,m,n} {}_{-2}Y_{\ell m}(\iota, \delta) A_{\ell m n} \exp [ i (\Omega_{\ell m n} t + \phi_{\ell m n} )] ,
\end{aligned}
\end{equation}
where the complex frequency is given by $\Omega_{\ell m n} = 2\pi f_{\ell m n} + i/\tau_{\ell m n}$. The term ${}_{-2}Y_{\ell m}(\iota,\delta)$ represents the spin-weighted spherical harmonics (of spin weight $-2$), which depend on the inclination angle $\iota$ between the \ac{BH}'s spin axis and the line of sight. The azimuth angle $\delta$ is set to zero, as it is degenerate with the mode phases. We use spherical harmonics as an approximation for the spheroidal harmonics, which provides the useful time-domain symmetry $h_{\ell m}(t) = (-1)^\ell h^*_{\ell -m}(t)$ (the asterisk denotes complex conjugation).
We have explicitly verified that this approximation is robust and does not alter our main conclusions. A detailed comparison with a general spheroidal model is provided in the Appendix \ref{appen:b}.

For the analysis of GW231028, we employ the $\fs$ method, a framework that enhances computational efficiency by analytically maximizing the likelihood function over the linear parameters of the model (the amplitudes and phases of the \acp{QNM}). The effectiveness and reliability of this approach for ringdown studies have been demonstrated in prior works \citep{Wang:2024jlz,Wang:2024yhb}. The specific implementation used here--including the preprocessing of strain data, the noise estimation technique, and the methods for reconstructing posterior samples and computing Bayesian evidence--is identical to the validated pipeline detailed in \citet{Wang:2025rvn}.

To further ensure the robustness of our results, we complement the primary $\fs$ analysis with two independent validation tests. First, we perform a parallel parameter estimation using the standard \ac{TTD} method to cross-check the mode identification. Second, we conduct an injection study using the \ac{NR} waveform SXS:BBH:1282, which closely mimics the properties of GW231028, to verify the recovery of the signal using our pipeline. Detailed configurations for these analyses are available in the Appendix \ref{appen:a}.

\begin{figure}
\centering
\includegraphics[width=0.5\textwidth,height=7cm]{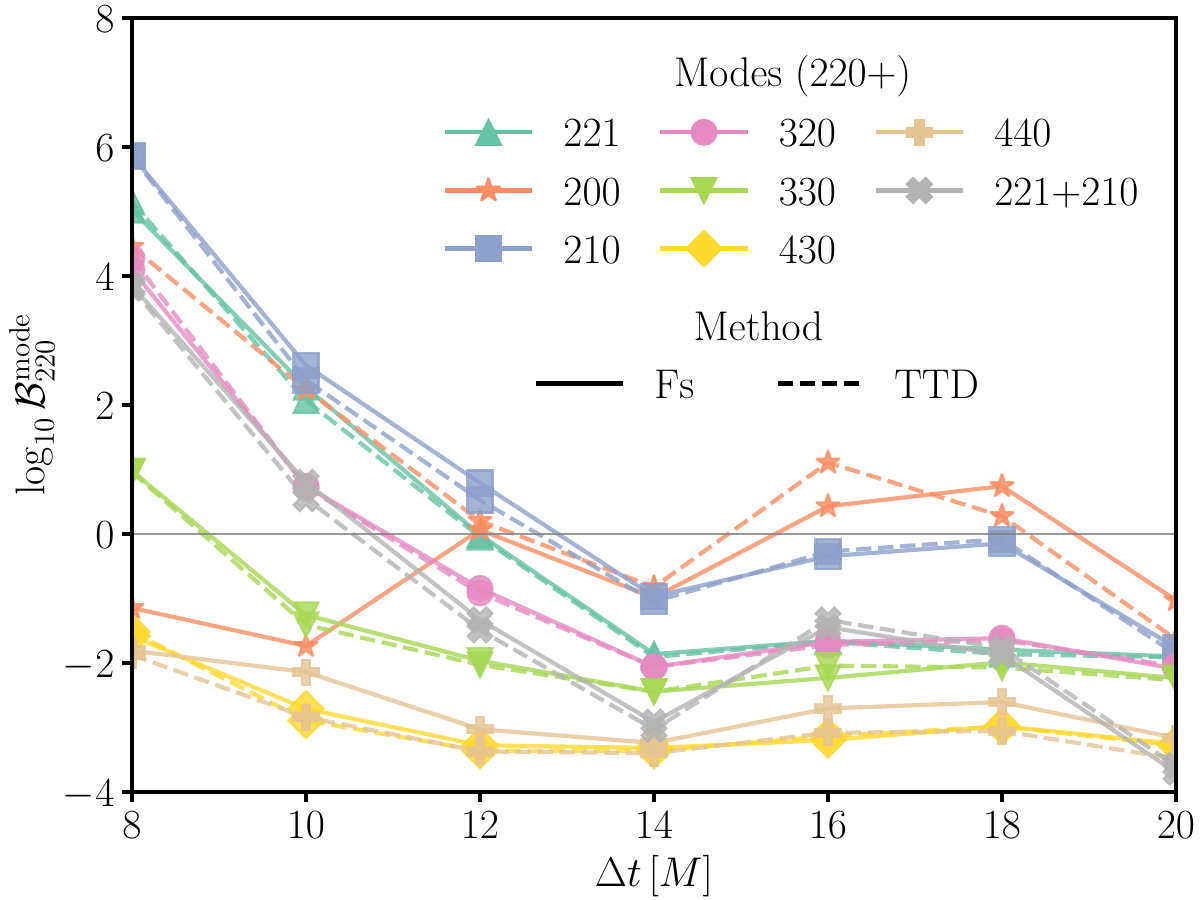}
\caption{
Bayesian evidence for various two-mode \ac{QNM} combinations in the ringdown analysis of GW231028. The vertical axis displays the $\log_{10}$(Bayes factor) for each combination relative to a model containing only the fundamental ($\ell=|m|=2, n=0$, or $220$) mode. The horizontal axis indicates the analysis start time ($\Delta t$), measured as a delay from the signal's polarization peak. Results are derived using the $\fs$ method and the \ac{TTD} method.
}\label{fig:bfs_m2}
\end{figure}

To perform the analysis, we first isolate the ringdown portion of the signal. 
The sky location and polarization angle \((\text{RA}, \text{DEC}, \psi)\) are fixed to their maximum-likelihood values of \((0.04, -0.10, 1.26)\), as obtained from the LVK Collaboration’s full \ac{IMR} analysis using the SEOBNRv5PHM waveform model \citep{Ramos-Buades:2023ehm}.
We acknowledge that GW231028 is subject to significant systematic uncertainties across different waveform models due to its high spin and unequal mass nature \citep{LIGOScientific:2025slb}. Consequently, fixing extrinsic parameters based on a single model could a priori introduce biases if the model imperfectly captures the pre-merger signal. However, establishing a fixed reference frame is a prerequisite for our ringdown analysis. To assess the reliability of our method given the signal's complexity, we performed a validation test using a \ac{NR} injection (SXS:BBH:1282) that mimics the source properties. As detailed in the Appendix \ref{appen:c}, this test confirms that our pipeline can robustly recover the ringdown modes from such a high-spin signal when the analysis is performed in a consistent reference frame.

We address the uncertainty in the start time of the linear ringdown by performing the analysis over a range of start times, $t_c$, from $\Delta t = [8, 20]\,M$ in steps of $2\,M$ after the signal's polarization peak, $t_c^{\mathrm{pol}}=\underset{t}{\max}|h_{+}-ih_{\times}|^2$. For GW231028, with a redshifted remnant mass of $M \approx 246.4\,M_\odot$ \citep{LIGOScientific:2025slb}, the peak time is $t_c^{\mathrm{pol}} = 1382542224.18917$ GPS.\footnote{In this context, $1\,M$ corresponds to approximately $1.2$ ms. For the fixed sky location, the polarization peak time at the LIGO Hanford detector is $1382542224.20303$ GPS.} The prior on the redshifted final mass is uniform over the range $[50, 300]\,M_\odot$. All other parameter priors are identical to those used in the analysis of \citet{Wang:2025rvn}.

The \ac{QNM} frequencies of the $\ell|m|n =210$ and $330$ modes in GW231028 lie close to $60$ Hz and $120$ Hz, respectively—frequencies at which the detector data exhibit prominent power lines. To avoid potential spectral contamination, we pre-processed the strain data using the LineCleaner method \citep{Farr:LineCleaner} to subtract these lines. However, as our ringdown analysis uses a $0.4$ s segment—sufficiently long to suppress spectral leakage effects \citep{Siegel:2024jqd}—the impact of these lines is expected to be small, and the subtraction primarily serves to enhance robustness.

\begin{figure*}
\centering
\begin{subfigure}[b]{0.48\linewidth}
\centering
\includegraphics[width=\textwidth,height=8cm]{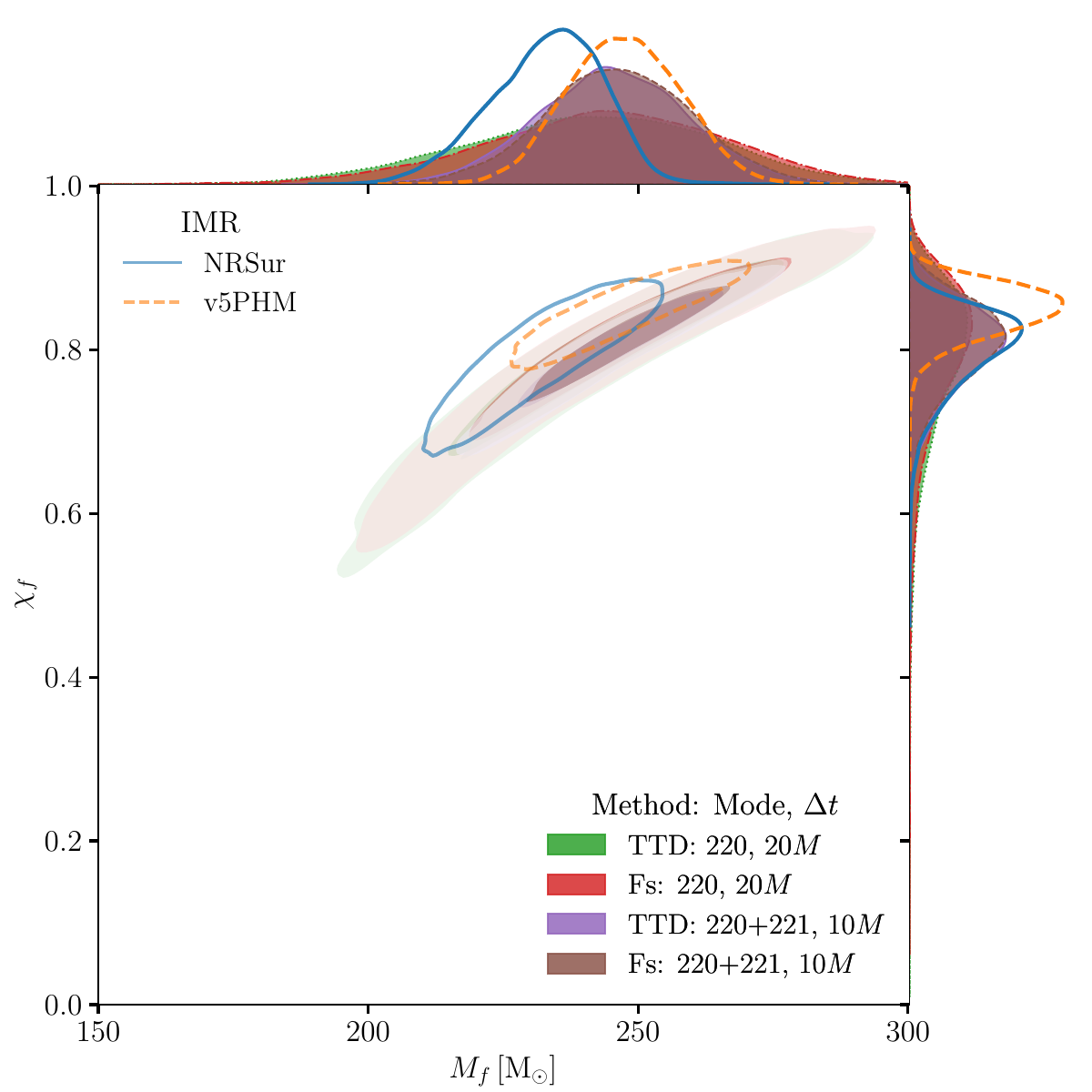}
\end{subfigure}%
\begin{subfigure}[b]{0.48\linewidth}
\centering
\includegraphics[width=\textwidth,height=8cm]{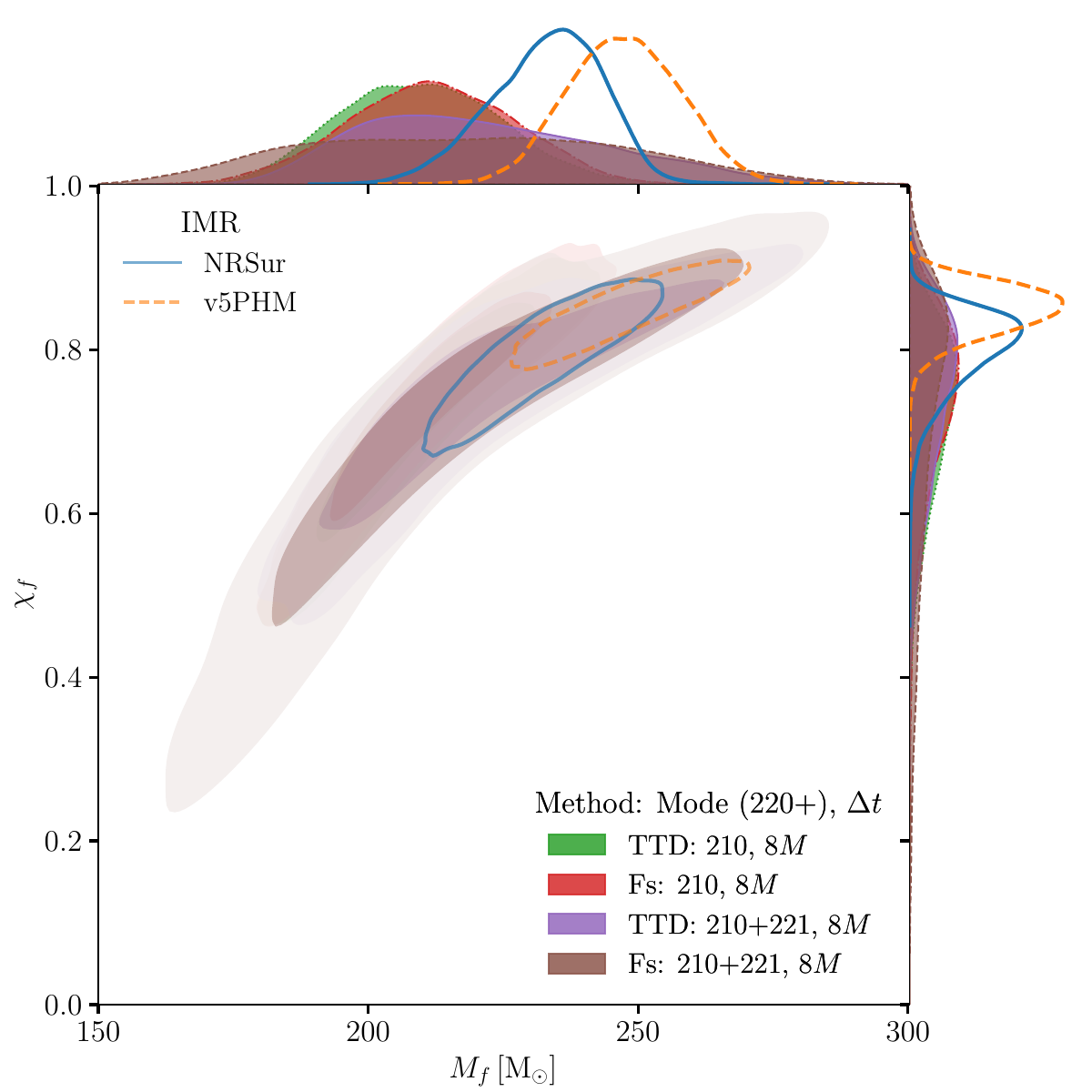}
\end{subfigure}%
\caption{
Posterior distributions for the redshifted final mass $M_f$ and final spin $\chi_f$ of the GW231028 remnant. The results from our ringdown analyses, using both the $\fs$ method and the \ac{TTD} method, are compared against the official LVK posteriors \citep{LIGOScientific:2025slb} derived from full \ac{IMR} analyses using the NRSur7dq4 and SEOBNRv5PHM waveform models.
For results of the full \ac{IMR} analyses, we show contours at $90\%$ credible level.
For results of the ringdown analyses, the contours enclose the $60\%$ and $90\%$ credible regions. The top and right side panels show the corresponding one-dimensional marginalized posterior distributions.
}\label{fig:fmfs_m2}
\end{figure*}

\section{Multimode search}\label{sec:bayes}
Our search for a multimode signal begins by looking for a sub-dominant mode accompanying the fundamental $(\ell|m|n = 220)$ mode, with candidates selected from the set $\ell|m|n \in \{221, 200, 210, 320, 330, 430, 440, 221+210\}$. The Bayes factors for each combination relative to the fundamental-mode-only model are computed across a range of start times, as shown in Fig.~\ref{fig:bfs_m2}.
Two candidate models stand out with significant statistical support: the first overtone combination ($220+221$) and the higher-order combination ($220+210$). To validate these findings, we performed the analysis using both the $\fs$ method and the full \ac{TTD} Bayesian parameter estimation. As detailed in Fig.~\ref{fig:fmfs_m2}, the two methods show excellent agreement for these modes, demonstrating that the detection is robust against sampler systematics.

\begin{figure*}
\centering
\includegraphics[width=0.88\textwidth,height=15cm]{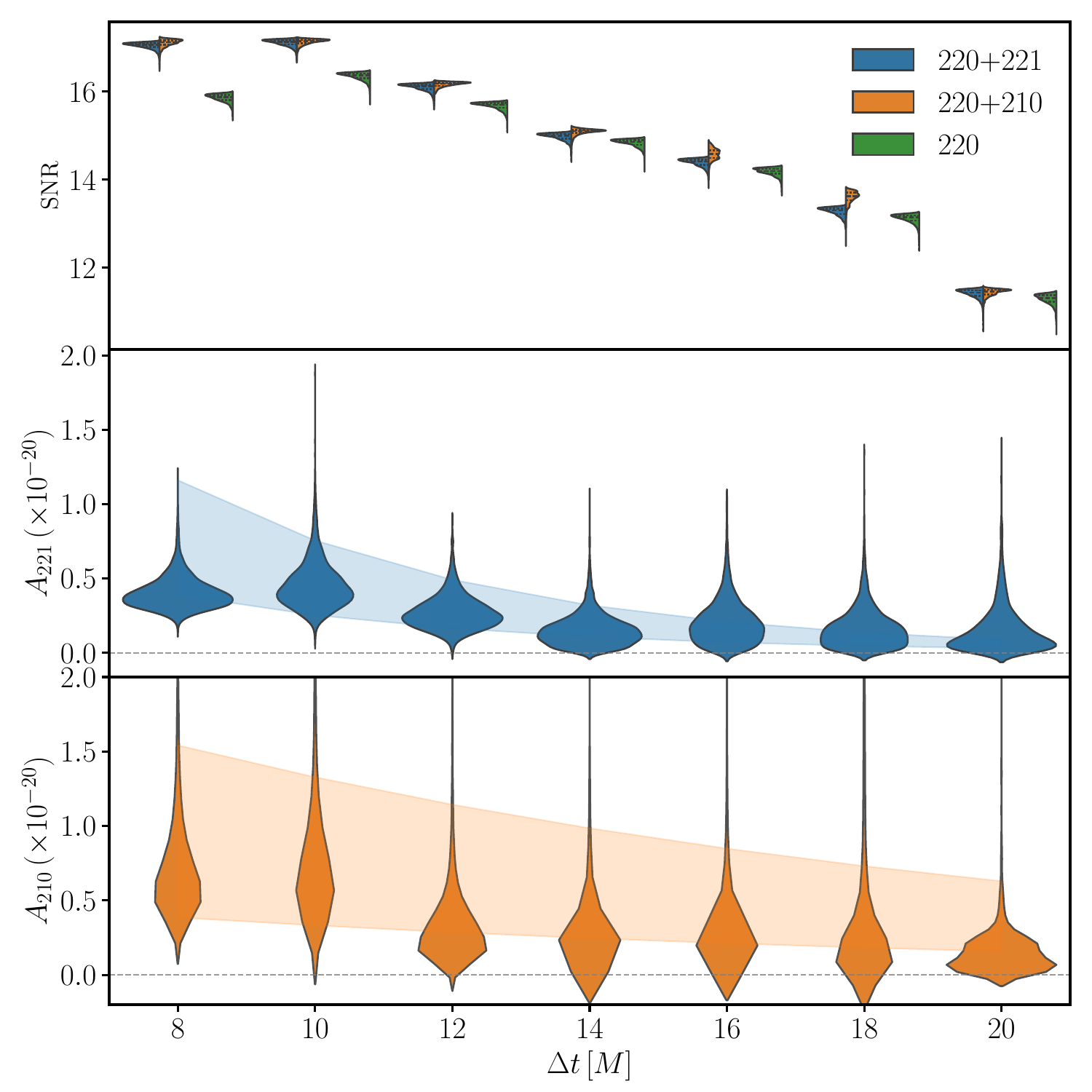}
\caption{
Results for the $220+221$ and $220+210$ mode analyses as a function of the analysis start time $\Delta t$ relative to the peak time.
Top panel: The posterior distributions (violin plots) of the \ac{SNR} for different combinations ($220$ in green, $220+221$ in blue, and $220+210$ in orange).
Middle and Bottom panels: Posterior distributions for the amplitudes of the first overtone mode ($A_{221}$) and the higher-order mode ($A_{210}$), defined at the corresponding start time $\Delta t$. 
To assess consistency, we include theoretical decay bands ($90\%$ credible level) shown in light blue for $A_{221}$ and light orange for $A_{210}$. These bands represent the expected amplitude evolution governed by the sub-dominant mode damping times, anchored to the reference posteriors recovered at $\Delta t = 10M$ for the $221$ mode and $\Delta t = 8M$ for the $210$ mode. The agreement between the recovered distributions and the predicted bands demonstrates the temporal consistency of the signals.
}\label{fig:snr_amps}
\end{figure*}

For the $220+210$ mode, we find decisive evidence at $\Delta t = 8\,M$, with $\log_{10} \mathcal{B} \approx 5.9$ using both $\fs$ and \ac{TTD} methods. Validation against the \ac{NR} injection SXS:BBH:1282 (see below and Appendix \ref{appen:c}) reveals that a sub-dominant $(\ell,|m|)=(2,1)$ mode with comparable amplitude is physically expected for a remnant with these source properties (high spin and mass ratio)\footnote{We thank the anonymous referee for their insightful comments, which prompted us to rigorously re-examine the physical significance of this mode and confirm its robust detection.}. 
For the $220+221$ mode, we recover strong evidence at $\Delta t = 8\,M$ and $10\,M$. While the statistical support is strongest at $\Delta t = 8\,M$ ($\log_{10} \mathcal{B} \approx 5.0$), we adopt the analysis starting at $\Delta t = 10\,M$ as our primary reference. This choice is driven by the physical consistency of the inferred remnant properties: as demonstrated in Fig.~\ref{fig:fmfs_m2} and further comparisons in the Appendix \ref{appen:b}, the joint posterior distributions for the final mass and spin derived from the $220+221$ model at this start time exhibit good alignment with the full \ac{IMR} predictions. At this start time, the evidence remains decisive\footnote{Here, we use a scale by Jeffrey \citep{jeffreys1998theory}, where a Bayes factor greater than $100$ is considered ``decisive" evidence.}, favoring the inclusion of the $221$ mode by a factor of $\sim 10^{2.3}$. The corresponding network \ac{SNR} for the two-mode signal at $\Delta t = 10\,M$ is approximately $17.1$.

However, at early times ($\Delta t = 8, 10\,M$), we find that neither the $220+221$ model nor the $220+210$ model provides constraints on the remnant mass and spin that fully encompass the \ac{IMR} predictions. This occurs because the contribution of the first overtone ($221$) is comparable to that of the $(2,1)$ angular mode and both are non-negligible (see Fig.~\ref{fig:snr_amps} and Table.~\ref{tab:results}).
This interpretation is rigorously corroborated by our three-mode analysis ($220+221+210$), which, by accounting for both components simultaneously, maintains much better consistency with \ac{IMR} expectations for $\Delta t\in[8,14]\,M$ (see Appendix \ref{appen:b}), beyond which the $221$ mode is too weak to be reliably measured and its inclusion may degrade parameter estimation. Nevertheless, due to the moderate \ac{SNR} of GW231028, the three-mode hypothesis incurs a statistical penalty for its expanded parameter space (Occam's razor), yielding lower Bayes factors than the two-mode hypotheses (Fig.~\ref{fig:bfs_m2}). Thus, while the current signal strength does not support a decisive simultaneous detection, the time-dependent behavior of the parameter constraints strongly implies the physical presence of both modes.

The amplitude of the $221$ mode is confidently constrained away from zero, with a significance of $7.2\sigma$ at $\Delta t=10\,M$ (using the $\fs$ method). The \ac{TTD} analysis yields an even higher significance of $16.5\sigma$, further reinforcing the detection. Here, the ``$p_A$" value quantifies how strongly the posterior distribution for the amplitude excludes zero, following the methodology for non-negative \ac{QNM} amplitudes outlined in the Supplemental Material of Ref.~\cite{LIGOScientific:2025obp}. The mode decays as expected, falling below $3\sigma$ credibility at $\Delta t=12\,M$.

The remnant parameters inferred from the $220+221+210$ model at $\Delta t\in [8,14]\,M$ are consistent with results from full \ac{IMR} analyses using both the NRSur7dq4 \citep{Varma:2019csw} and SEOBNRv5PHM \citep{Ramos-Buades:2023ehm} waveforms (Fig.~\ref{fig:fmfs_m2}). The consistency with NRSur7dq4 is particularly notable given the construction of our analysis window. We define the ringdown reference time based explicitly on the SEOBNRv5PHM model, identifying the waveform peak time associated with the maximum likelihood parameters (including sky position and geocent time) from the LVK posterior. Despite this specific reliance on SEOBNRv5PHM for temporal alignment, the resulting ringdown constraints overlap significantly with the independent predictions of NRSur7dq4. This cross-model agreement underscores the robustness of our reference time definition and the reliability of the recovered remnant properties. 

Specifically, from the $220+221$ ($220+210$) model at $\Delta t = 10\,M$ ($\Delta t=8 \,M$), the $\fs$ method constrains the redshifted final mass to be $246.2_{-22.4}^{+22.3}$ ($211.6_{-25.6}^{+25.0}$) $M_\odot$ and the final spin to be $0.81_{-0.10}^{+0.07}$ ($0.75_{-0.23}^{+0.13}$) at the $90\%$ credible level. In contrast, a single-mode analysis fails to produce results consistent with the \ac{IMR} posteriors until much later start times ($\Delta t \geq 16\,M$), where the lower \ac{SNR} degrades the constraints.
Other two-mode combinations, such as $220+320$, are disfavored by Bayes factors and fail to yeild physically meaningful posteriors.

\begin{table}[htbp]
\centering
\caption{Parameter estimation results and model selection statistics. The columns display the analysis method (Fs or TTD), the waveform model with the start time $\Delta t$ relative to the peak, the final mass $M_f$, the final dimensionless spin $\chi_f$, the matched filter \ac{SNR} $\rho$, the $\log_{10}$ Bayes factor favoring the hypothesis with the higher-order mode, and the significance level $p_A$ (see the Supplemental Material of Ref.~\citep{LIGOScientific:2025obp} for its definition) of the higher-order mode amplitude deviation from zero.}
\label{tab:results}
\resizebox{\columnwidth}{!}{%
\begin{tabular}{l|cr|ccccc}
\toprule
& Model & $\Delta t$ & $M_f$ & $\chi_f$ & $\rho$ & $\log_{10}\mathcal{B}$ & $p_A$  \\ {}&(220+)&\\
\hline
\multirow{4}{*}{Fs} 
	   & 221 & 8M  & $266.0^{+16.8}_{-17.8}$ & $0.87^{+0.04}_{-0.06}$ & $17.1^{+0.1}_{-0.2}$ & 5.0 & 15.0 \\
	   & 221 & 10M & $246.2^{+22.3}_{-22.4}$ & $0.81^{+0.07}_{-0.10}$ & $17.1^{+0.1}_{-0.2}$ & 2.3 & 7.2 \\
	   & 210 & 8M  & $211.6^{+25.0}_{-25.6}$ & $0.75^{+0.13}_{-0.23}$ & $17.1^{+0.1}_{-0.2}$ & 5.9 & 2.8 \\
	   & 210 & 10M & $192.8^{+33.4}_{-26.8}$ & $0.61^{+0.23}_{-0.37}$ & $17.1^{+0.1}_{-0.2}$ & 2.6 & 1.9 \\
\hline
\multirow{4}{*}{TTD} 
	   & 221 & 8M  & $265.4^{+16.0}_{-17.7}$ & $0.87^{+0.04}_{-0.06}$ & $17.0^{+1.7}_{-1.7}$ & 5.1 & 8.8 \\
	   & 221 & 10M & $245.2^{+23.1}_{-22.8}$ & $0.81^{+0.07}_{-0.11}$ & $17.1^{+1.6}_{-1.6}$ & 2.1 & 16.5 \\
	   & 210 & 8M  & $209.4^{+25.0}_{-23.0}$ & $0.73^{+0.13}_{-0.21}$ & $17.0^{+1.7}_{-1.7}$ & 5.9 & 2.5 \\
	   & 210 & 10M & $194.7^{+30.1}_{-22.5}$ & $0.62^{+0.21}_{-0.28}$ & $17.0^{+1.7}_{-1.7}$ & 2.4 & 1.4 \\
\hline
\end{tabular}%
}
\end{table}

To rigorously validate that the high Bayes factors for $221$ and $210$ modes arise from physical signal content, we performed a parallel analysis on the \ac{NR} injection SXS:BBH:1282. This waveform models a high-spin, mass-ratio $q \approx 2$ system that closely mimics the source properties of GW231028 (see Appendix \ref{appen:c} for full injection parameters and detailed results). The injection analysis confirms that for such a system, the $(2,1)$ mode is a significant sub-dominant feature (amplitude $\sim 1/3$ of the fundamental mode). Crucially, the test also faithfully reproduces the contribution of the first overtone ($221$ mode) and the higher-order ($(2,1)$ mode). Our pipeline successfully recovers both sub-dominant components, yielding Bayes factors and parameter evolutions that closely mirror the multimode signature observed in the real event.

\begin{figure}
\centering
\includegraphics[width=0.5\textwidth,height=8cm]{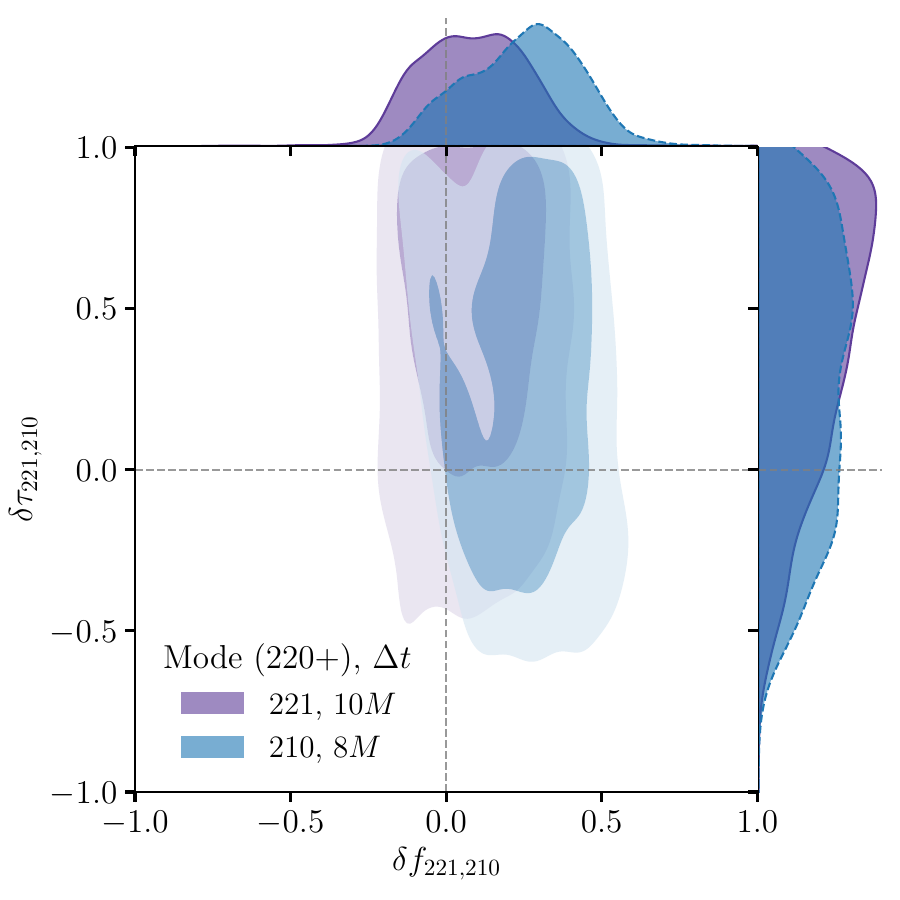}
\caption{
Results of testing the no-hair theorem using the ringdown signal of GW231028, analyzed starting from $10\,M$ ($8\,M$) after the strain peak with $\fs$ method considering $220+221$ ($220+210$) mode combinations. The horizontal axis represents the deviation \(\delta f_{221,210}\) of the \(f_{221,210}\), while the vertical axis denotes the deviation \(\delta \tau_{221,210}\) of the \(\tau_{221,210}\). The deviations are formulated as \(f_{221,210} \times (1 + \delta f_{221,210})\) and \(\tau_{221,210} \times (1 + \delta \tau_{221,210})\), respectively.
The contours enclose the $60\%$ and $90\%$ credible regions. The top and right side panels show the corresponding one-dimensional marginalized posterior distributions.
}\label{fig:dfdt221}
\end{figure}

Building on the confident detection of these sub-dominant modes, we perform a test of the no-hair theorem. We parameterize potential fractional deviations in the frequency ($\delta f$) and damping time ($\delta \tau$) from the \ac{GR} predictions. As shown in Fig.~\ref{fig:dfdt221}, the results are entirely consistent with \ac{GR}. For the $221$ mode at $\Delta t = 10\,M$, the constraints given by the $\fs$ method are $\delta f_{221} = 0.09_{-0.25}^{+0.26}$ and $\delta \tau_{221} = 0.48_{-0.82}^{+0.48}$ ($90\%$ credible level). Similarly, the $210$ mode at $\Delta t = 8\,M$ shows no deviation. We also computed the Bayes factors comparing the hypothesis allowing for deviations against the strict GR hypothesis. Using the $\fs$ method, we found $\log_{10}\mathcal{B}^{\text{non-GR}}_{\text{GR}} \approx -0.7$ for the $220+221$ combination and $\log_{10}\mathcal{B}^{\text{non-GR}}_{\text{GR}} \approx -0.5$ for the $220+210$ combination. 
These negative values indicate a statistical preference for the simpler GR model, as the model with deviations is penalized for its larger prior volume while providing no significant improvement in fit, consistent with the validity of the no-hair theorem.

\section{Discussion and Conclusions}\label{sec:con}
In this work, we present decisive evidence for multimode content in the ringdown signal of a \ac{BBH} merger, specifically identifying both the first overtone ($\ell|m|n = 221$) and the higher-order ($\ell|m|n = 210$). Our analysis of GW231028 reveals a statistically significant detection of the $221$ mode at $\Delta t = 10\,M$, supported by a Bayes factor of $\sim 10^{2.3}$ and a nonzero amplitude confirmed at $7.2\sigma$ ($16.5\sigma$) credibility using the $\fs$ (\ac{TTD}) method. Furthermore, we find robust evidence for the $(2,1)$ harmonic at $\Delta t = 8\,M$, with Bayes factors reaching $\sim 10^{5.9}$. Crucially, these detections are corroborated by parallel analyses using the full \ac{TTD} sampling and validated against the \ac{NR} injection SXS:BBH:1282. The injection study confirms that for a remnant with the high spin and mass ratio inferred for GW231028, the coexistence of a strong early-time overtone and a sub-dominant $(2,1)$ mode is a physically expected signature. 
Representing a robust identification of an overtone or a sub-dominant mode in a moderate \ac{SNR} event, this result---appearing as an independent, contemporaneous discovery alongside the high-\ac{SNR} detection in GW250114 \citep{LIGOScientific:2025obp}---demonstrates the observable ubiquity of these spectral features. Furthermore, while GW250114 represents an idealized non-precessing case, our work confirms these multimode signatures in a precessing, unequal-mass system. Consequently, this work marks a significant step forward in the field of \ac{BH} spectroscopy, extending the reach of precision tests of \ac{GR} beyond the loudest ``golden" events to the broader population of complex binary mergers.

While the current \ac{SNR} of the ringdown signal statistically favors two-mode hypotheses over a simultaneous three-mode model, the physical consistency of our findings is reinforced by the full \ac{IMR} constraints. We show that including the $221$ mode is essential for reconciling the ringdown-derived mass and spin with predictions from two independent waveform models, NRSur7dq4 and SEOBNRv5PHM. This correction eliminates the systematic biases present when only the fundamental mode or the $(2,1)$ harmonic are considered in isolation. Moreover, the consistency with NRSur7dq4 is particularly notable given that our analysis was anchored to the peak time defined by only the SEOBNRv5PHM model. 
While our agnostic analysis robustly identifies the presence of a second mode, the specific identification of this mode is subject to the inherent limitations of black-hole spectroscopy. Due to the non-orthogonality of QNMs, multiple mode combinations can reproduce the signal with comparable fidelity \citep{CalderonBustillo:2020rmh, Chandra:2025ipu}. In particular, sub-dominant modes (such as $(2,0)$ or $(3,2)$) that are spectrally close to dominant modes (like $(2,1)$ or $(2,2)$) can sometimes be statistically favored by the Bayesian evidence purely due to their ability to fit residuals, even if they are not the physically dominant excitation. This effect, combined with the Occam penalty against adding additional modes to restore consistency with IMR predictions, limits the capability of agnostic spectroscopy to provide strong tests of the no-hair theorem without external constraints or validation against numerical relativity counterparts.

The astrophysical implications of these findings are particularly compelling when contrasted with the recent analysis of GW231123 \citep{Wang:2025rvn}. In that event, the ringdown was dominated by the $\ell|m|n = 200$ mode. For GW231028, however, we find no evidence for the $200$ mode. This contrast reinforces the view that agnostic spectroscopy, by relaxing the tight priors of quasi-circular \ac{IMR} models, can robustly identify diverse spectral features—whether they are the $200$ mode or the overtones. Instead, the clear identification of the $221$ overtone alongside the $(2,1)$ harmonic paints a coherent picture of a remnant formed from a quasi-circular merger following a prolonged inspiral. 
This distinction highlights the powerful potential of the ringdown analysis: it not only provides a platform for testing \ac{GR} and the no-hair theorem but also serves as a largely independent and complementary probe of the final state, offering a new observational tool to distinguish between different astrophysical formation channels.

The successful and distinct multimode detections in both GW231028 and GW231123 signal that \ac{BH} spectroscopy is transitioning from a theoretical goal to a practical tool in \ac{GW} astronomy. As detector sensitivity continues to improve, such analyses will become increasingly common, enabling precision tests of strong-field gravity and offering unprecedented insights into the diverse universe of \ac{BH} mergers.

This work made use of the following software packages: \texttt{astropy} \citep{astropy:2013, astropy:2018, astropy:2022}, \texttt{matplotlib} \citep{Hunter:2007}, \texttt{numpy} \citep{numpy}, \texttt{python} \citep{python}, \texttt{scipy} \citep{2020SciPy-NMeth, scipy_12522488}, \texttt{Bilby} \citep{Ashton:2018jfp, Bilby_2602178}, \texttt{PyCBC} \citep{2012PhRvD..85l2006A}, \texttt{corner} \citep{corner-Foreman-Mackey-2016, corner.py_14209694}, \texttt{h5py} \citep{collette_python_hdf5_2014, h5py_7560547}, \texttt{pandas} \citep{mckinney-proc-scipy-2010, pandas_13819579}, \texttt{Jupyter} \citep{2007CSE.....9c..21P, kluyver2016jupyter}, \texttt{qnm} \citep{Stein:2019mop}, \texttt{lalsuite} \citep{lalsuite} \texttt{swiglal} \citep{swiglal}, \texttt{GWpy} \citep{gwpy,Macleod:2021goi}, \texttt{pykerr} \citep{collin_capano_2023_10056495,Leaver:1985ax,Berti:2005ys}, and \texttt{sxs} \citep{Boyle_The_sxs_package_2025}.

\begin{acknowledgements}
We thank the anonymous referee for their insightful comments.
This work was supported by ``the Natural Science Foundation of Liaoning Province" (Grant No. ZX20250217) and ``the Fundamental Research Funds for the Central Universities" at Dalian University of Technology.
This research has made use of data or software obtained from the Gravitational Wave Open Science Center~\cite{gwosc-url}, a service of LIGO Laboratory, the LIGO Scientific Collaboration, the Virgo Collaboration, and KAGRA~\cite{KAGRA:2023pio}. 
LIGO Laboratory and Advanced LIGO are funded by the United States National Science Foundation (NSF) as well as the Science and Technology Facilities Council (STFC) of the United Kingdom, the Max-Planck-Society (MPS), and the State of Niedersachsen/Germany for support of the construction of Advanced LIGO and construction and operation of the GEO600 detector. 
Additional support for Advanced LIGO was provided by the Australian Research Council. 
Virgo is funded, through the European Gravitational Observatory (EGO), by the French Centre National de Recherche Scientifique (CNRS), the Italian Istituto Nazionale di Fisica Nucleare (INFN) and the Dutch Nikhef, with contributions by institutions from Belgium, Germany, Greece, Hungary, Ireland, Japan, Monaco, Poland, Portugal, Spain.
KAGRA is supported by Ministry of Education, Culture, Sports, Science and Technology (MEXT), Japan Society for the Promotion of Science (JSPS) in Japan; National Research Foundation (NRF) and Ministry of Science and ICT (MSIT) in Korea; Academia Sinica (AS) and National Science and Technology Council (NSTC) in Taiwan of China.
\end{acknowledgements}

\appendix

\setcounter{equation}{0}
\setcounter{figure}{0}
\setcounter{table}{0}
\renewcommand{\theequation}{S\arabic{equation}}
\renewcommand{\thefigure}{S\arabic{figure}}
\renewcommand{\thetable}{S\arabic{table}}

\section{Supplemental Methods: the $\mathcal{F}$-statistic}\label{appen:a}

\subsection{Analysis Frameworks for Ringdown Signals}

This appendix outlines the $\fs$ framework used to analyze the ringdown phase of gravitational wave events. 

\subsubsection{Core Data Model and Noise Treatment}

The detector output $d(t)$ is modeled as the sum of a gravitational wave signal $h(t,\Theta)$ and stationary, Gaussian noise $n(t)$. To focus on the post-merger ringdown, the analysis uses a truncated segment of the full data stream, a standard practice established in prior work \citep{Isi:2021iql,Wang:2023mst,Wang:2024liy,Siegel:2024jqd}.

The log-likelihood function for Gaussian noise is defined as:
\begin{equation}
\begin{aligned}\label{eq:ll0}
\ln\mathcal{L}(\Theta) &= -\frac{1}{2}\langle \mathbf{d}(t)-\mathbf{h}(t,\Theta)|\mathbf{d}(t)-\mathbf{h}(t,\Theta)\rangle+C_0 \\
&= \ln\Lambda(\Theta)-\frac{1}{2}\langle \mathbf{d}|\mathbf{d}\rangle+C_0,
\end{aligned}
\end{equation}
where $C_0$ is a normalization constant, and the key quantity is the log-likelihood ratio:
\begin{equation}\label{eq:ll1}
	\ln\Lambda(\Theta)=\langle \mathbf{d}|\mathbf{h}(t,\Theta)\rangle-\frac{1}{2}\langle \mathbf{h}(t,\Theta)|\mathbf{h}(t,\Theta)\rangle.
\end{equation}
The inner product $\langle \cdot | \cdot \rangle$ incorporates the noise characteristics. For discrete data, it is computed as:
\begin{equation}
	\langle \mathbf{h}_1(t) | \mathbf{h}_2(t)\rangle = \mathbf{h}_1(t)\mathcal{C}^{-1}\mathbf{h}_2^{\intercal}(t),
\end{equation}
where $\mathcal{C}$ is the noise covariance matrix and $\intercal$ denotes the transpose.
We employ the log-likelihood in \eqref{eq:ll0} for the \ac{TTD} method.

Constructing a reliable $\mathcal{C}$ is important. It is derived from the noise \ac{ACF}, which itself is estimated from a carefully processed version of the raw data. The processing pipeline, following \citep{Wang:2023mst,Wang:2024liy}, involves:
\begin{enumerate}
    \item Downsampling from $4096$ Hz to $2048$ Hz using a Butterworth filter.
    \item Applying a high-pass filter at $20$ Hz with a Finite Impulse Response filter \citep{KhanFIR2020}.
	\item Estimating the one-sided \ac{PSD} using the Welch method \citep{1967D.Welch} combined with an inverse spectrum truncation algorithm, implemented via the PyCBC package (v2.7.2) \citep{2012PhRvD..85l2006A}.
	\item Obtaining the \ac{ACF} by inverse Fourier transforming the \ac{PSD} and constructing the Toeplitz matrix $\mathcal{C}$.
\end{enumerate}

This noise model has been rigorously validated, showing consistency with full inspiral-merger-ringdown analyses and across different sampling rates \citep{Wang:2023mst, Wang:2024liy}, confirming its robustness against potential biases. For the final analysis, a $0.4$ s segment of the validated data is used.

\subsubsection{The $\mathcal{F}$-Statistic for Efficient Sampling}

The conventional time-domain ringdown analysis performs a direct, stochastic exploration of the \emph{entire} parameter space $\Theta$. This includes all parameters defining the ringdown signal model (e.g., mode amplitudes, phases, final mass, spin, inclination, start time). The sampler evaluates the full likelihood function (\eqref{eq:ll0}) at each proposed point in $\Theta$, allowing it to infer the complete posterior probability distribution for all parameters simultaneously. 

The $\fs$ method used in this study, is designed for greater computational efficiency by analytically handling a subset of the parameters. The core idea is to partition the model parameters into two groups:
\begin{itemize}
    \item Linear Parameters ($B^\mu$): These are parameters that enter the signal model $h(t)$ linearly. For ringdown modes, these correspond to the Cartesian components of the complex mode amplitudes (i.e., $A_{\ell mn}\cos\phi_{\ell mn}$ and $A_{\ell mn}\sin\phi_{\ell mn}$).
    \item Non-linear Parameters ($\bm\theta$): These parameters (e.g., final mass $M_f$, final spin $\chi_f$, inclination $\iota$, start time $t_c$, sky position, polarization angle $\psi$) enter the model non-linearly and still require stochastic sampling.
\end{itemize}

The signal is reformulated as a linear combination of basis waveforms $\mathbf{g}_\mu(t)$, which depend \emph{only} on the non-linear parameters $\bm\theta$:
$$\mathbf{h}(t) = B^\mu \mathbf{g}_\mu(t),$$
where the Einstein summation convention is used. For a mode $(\ell, m, n)$, the basis waveforms are:
\begin{equation}
\begin{aligned}
\mathbf{g}_{\ell mn,1} = &\left[F^+\cos(2\pi f_{\ell mn}t)+F^{\times}\sin(2\pi f_{\ell mn}t)\right] \\
						 &\times {}_{-2}Y_{\ell m}(\iota,\delta)\exp(-t/\tau_{\ell mn}), \\
\mathbf{g}_{\ell mn,2} = &\left[-F^{+}\sin(2\pi f_{\ell mn}t)+F^{\times}\cos(2\pi f_{\ell mn}t)\right] \\
						 &\times {}_{-2}Y_{\ell m}(\iota,\delta)\exp(-t/\tau_{\ell mn}),
\end{aligned}
\end{equation}
and the coefficients are $B^{\ell mn,1} = A_{\ell mn}\cos\phi_{\ell mn}$, $B^{\ell mn,2} = A_{\ell mn}\sin\phi_{\ell mn}$.

Substituting this into the log-likelihood ratio (\eqref{eq:ll1}) yields a quadratic form in $B^\mu$:
\begin{equation}\label{eq:ll2}
	\ln\Lambda(\Theta)=B^{\mu}\mathbf{s}_{\mu}(\bm\theta)-\frac{1}{2}B^{\mu}\mathbf{M}_{\mu\nu}(\bm\theta)B^{\nu},
\end{equation}
where $\mathbf{s}_{\mu} = \langle d|\mathbf{g}_{\mu}\rangle$ and $\mathbf{M}_{\mu\nu} = \langle \mathbf{g}_{\mu}|\mathbf{g}_{\nu}\rangle$. For any fixed $\bm\theta$, this likelihood is maximized analytically by solving $\partial\ln\Lambda / \partial B^{\nu} = 0$, giving the best-fit linear parameters:
\begin{equation}\label{eq:B}
\hat{B}^{\mu}=\mathbf{M}^{\mu\nu}\mathbf{s}_{\nu}.
\end{equation}

Plugging $\hat{B}^{\mu}$ back into \eqref{eq:ll2} produces the profile likelihood, which depends \emph{only} on $\bm\theta$:
\begin{align}\label{eq:fs1}
\ln\Lambda(\bm\theta) \equiv \mathcal{F} = \frac{1}{2}\mathbf{s}_{\mu}\mathbf{M}^{\mu\nu}\mathbf{s}_{\nu}.
\end{align}
This $\fs$ statistic is used for the stochastic sampling of the reduced parameter space $\bm\theta$, significantly speeding up the computation.

\subsection{Reconstructing Linear Parameter Posteriors}

While the $\fs$ method efficiently samples $\bm\theta$, the posterior distributions for the linear parameters $B^\mu$ (and thus the physical amplitudes $A_{\ell mn}$ and phases $\phi_{\ell mn}$) are not directly obtained. They are reconstructed in a post-processing step. For each sampled point $\bm\theta_i$, the conditional posterior for $B^\mu$ is a multivariate Gaussian:
\begin{equation}
	P(B^\mu | \boldsymbol{\theta}_i, \mathbf{d}) \sim \mathcal{N}\left(\hat{B}^\mu(\boldsymbol{\theta}_i), \mathbf{M}^{\mu\nu}(\boldsymbol{\theta}_i)\right).
\end{equation}
The full marginal posterior $P(B^\mu | \mathbf{d})$ is built by drawing samples from this Gaussian for each $\bm\theta_i$ sample.

The reconstruction naturally assumes a uniform prior on the Cartesian parameters $B^\mu$. However, this corresponds to an unphysical prior ($p(A_{\ell mn}) \propto 1/A_{\ell mn}$) on the amplitude $A_{\ell mn}$, while keeping the phase $\phi_{\ell mn}$ uniform (which is desirable).

To report results under physically motivated priors, importance resampling is applied. The weight for each reconstructed sample is:
\begin{equation}\label{eq:iw1}
w_i = \frac{p(B^{\mu}_i | \boldsymbol{\theta}_i, \otimes)}{p(B^{\mu}_i | \boldsymbol{\theta}_i,\ominus)},
\end{equation}
where $\ominus$ is the uniform sampling prior and $\otimes$ is the target prior. The target priors used in this study are: uniform amplitudes $A_{\ell mn} \in [0, 5\times 10^{-19}]$, flat $M_f\in[50, 300]\,\Msun$, flat $\chi_f\in[0, 0.99]$, flat $\cos\iota\in[-1,1]$, and uniform phases $\phi_{\ell mn} \in [0,2\pi]$.

\subsection{Calculating Evidence and Bayes Factors}

Model comparison via the Bayes factor $\mathcal{B}$ requires the full Bayesian evidence $\mathcal{Z}(\Theta)$. While nested sampling in the conventional Bayesian inference computes this directly, the $\fs$ method only yields the partial evidence $\mathcal{Z}(\bm\theta)$ for the non-linear parameters.

A novel formulation is used to reconstruct the full evidence $\mathcal{Z}(\Theta|\otimes)$ under the target prior $\otimes$:
\begin{equation}
\mathcal{Z}(\Theta|\otimes) \approx \frac{\mathcal{Z}(\bm\theta)}{N(\bm\theta)} \sum_{i=1}^{N(\bm\theta)} f(\bm\theta_i),\label{eq:z1}
\end{equation}
where the weighting function $f(\bm\theta_i)$ accounts for the evidence contribution of the linear parameters under the target prior, computed via importance sampling from the uniform prior $\ominus$ used in reconstruction:
\begin{align}
f(\bm\theta) &\approx \frac{\mathcal{Z}(B^{\mu}|{\bm\theta}, \ominus)}{N(B^{\mu}|{\bm\theta}, \ominus)} \sum_{j=1}^{N(B^{\mu}|{\bm\theta}, \ominus)} \frac{p(B^{\mu}_j | {\bm\theta}, \otimes)}{p(B^{\mu}_j | {\bm\theta}, \ominus)}.\label{eq:f1}
\end{align}
Here, $100\times N_0^2$ samples are drawn per $\bm\theta_i$ (with $N_0$ being the number of modes). This allows calculation of $\mathcal{B}^{\rm w1}_{\rm w2} = \mathcal{Z}_{\rm w1} / \mathcal{Z}_{\rm w2}$ for comparing different waveform models.

\subsection{Computational Implementation}

\begin{figure*}
\centering
\includegraphics[width=0.88\textwidth,height=12cm]{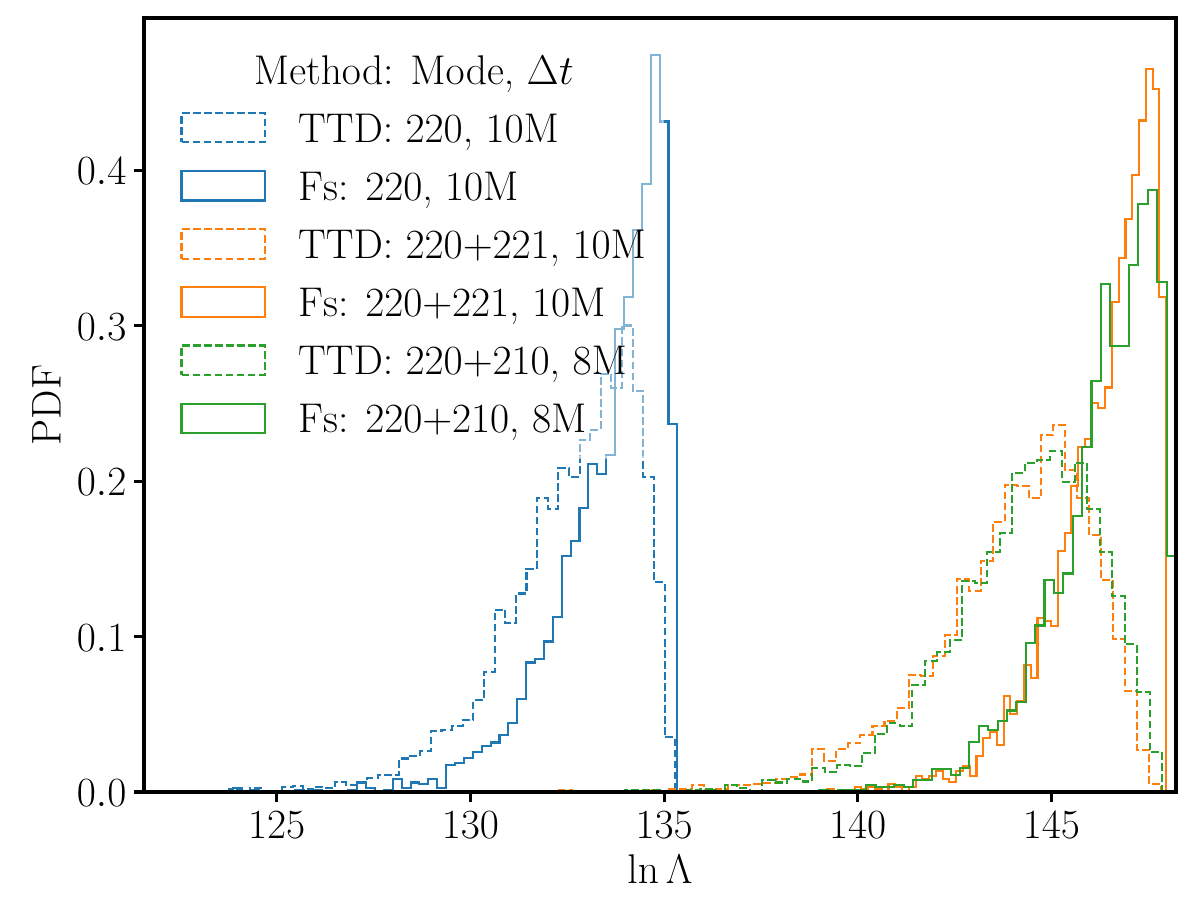}
\caption{
Posterior distributions of the log-likelihoods ratio ($\ln \Lambda$) for the ringdown analysis of GW231028.
The distributions are shown for the fundamental mode only ($220$, blue), the fundamental plus first overtone ($220+221$, orange), and the fundamental plus higher harmonic ($220+210$, green).
Solid lines represent results obtained using the $\fs$ method, while dashed lines correspond to the \ac{TTD} method.
The significant separation between the multi-mode distributions and the single-mode distribution illustrates the substantial \ac{SNR} gain and statistical preference for the multi-mode hypotheses.
Furthermore, the narrower and slightly shifted distributions for the $\fs$ method demonstrate its superior ability to locate the global likelihood maximum compared to the standard \ac{TTD} sampler.
}\label{fig:lls}
\end{figure*}

All analyses were performed using the {\sc Bilby} package~\citep[v2.4.0;][]{Ashton:2018jfp} with the nested sampler {\sc Dynesty} \citep[v2.1.5;][]{Romero-Shaw:2020owr}. The configuration used $1000$ live points, the ``rwalk" sampling method with $20$ autocorrelation lengths, a mix of ``diff" and ``volumetric" proposal methods, and the ``live-multi" point selection. Analyses were parallelized across $20$ CPU threads for efficiency.

\begin{figure*}
\centering
\begin{subfigure}[b]{0.45\linewidth}
\centering
\includegraphics[width=\textwidth,height=7.5cm]{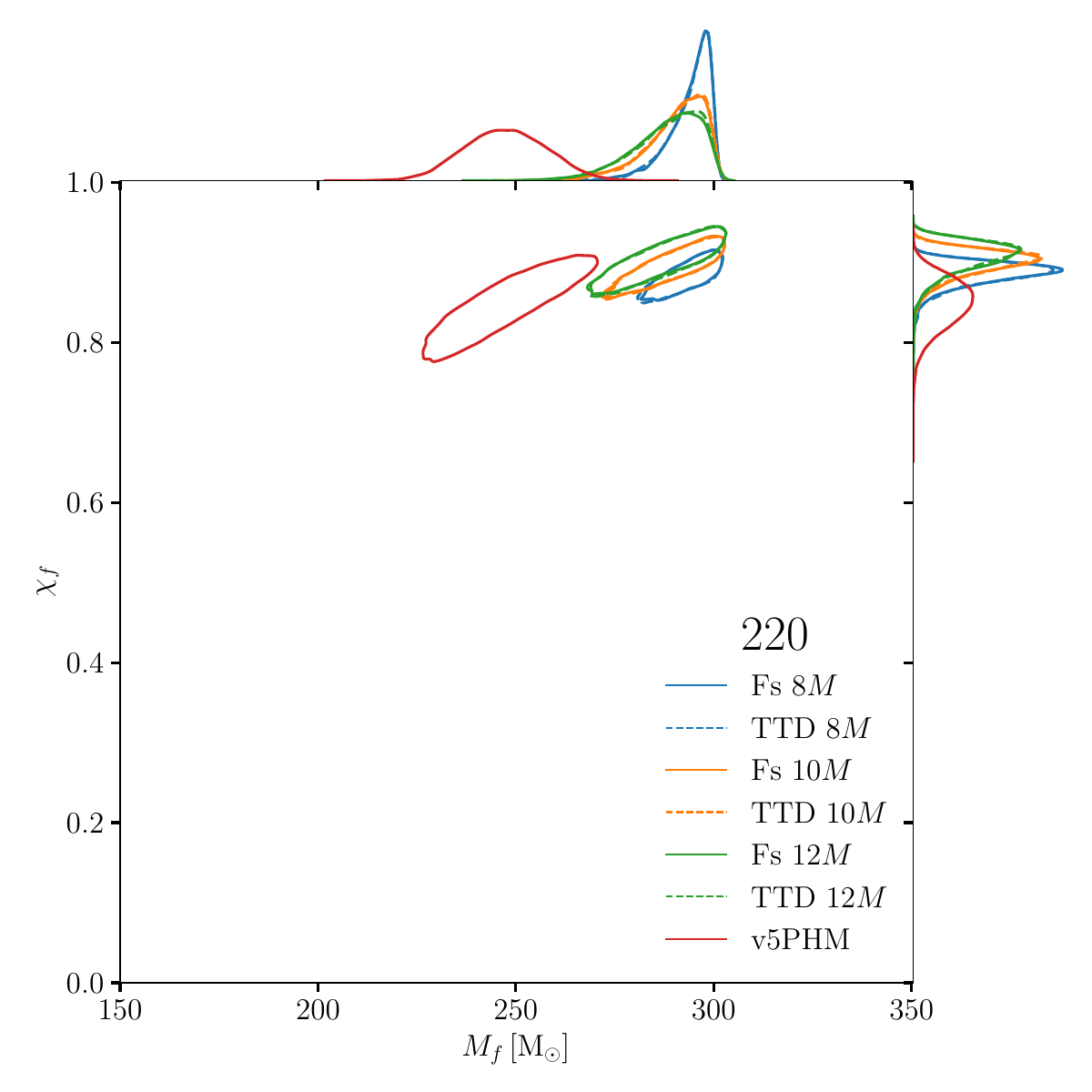}
\end{subfigure}%
\begin{subfigure}[b]{0.45\linewidth}
\centering
\includegraphics[width=\textwidth,height=7.5cm]{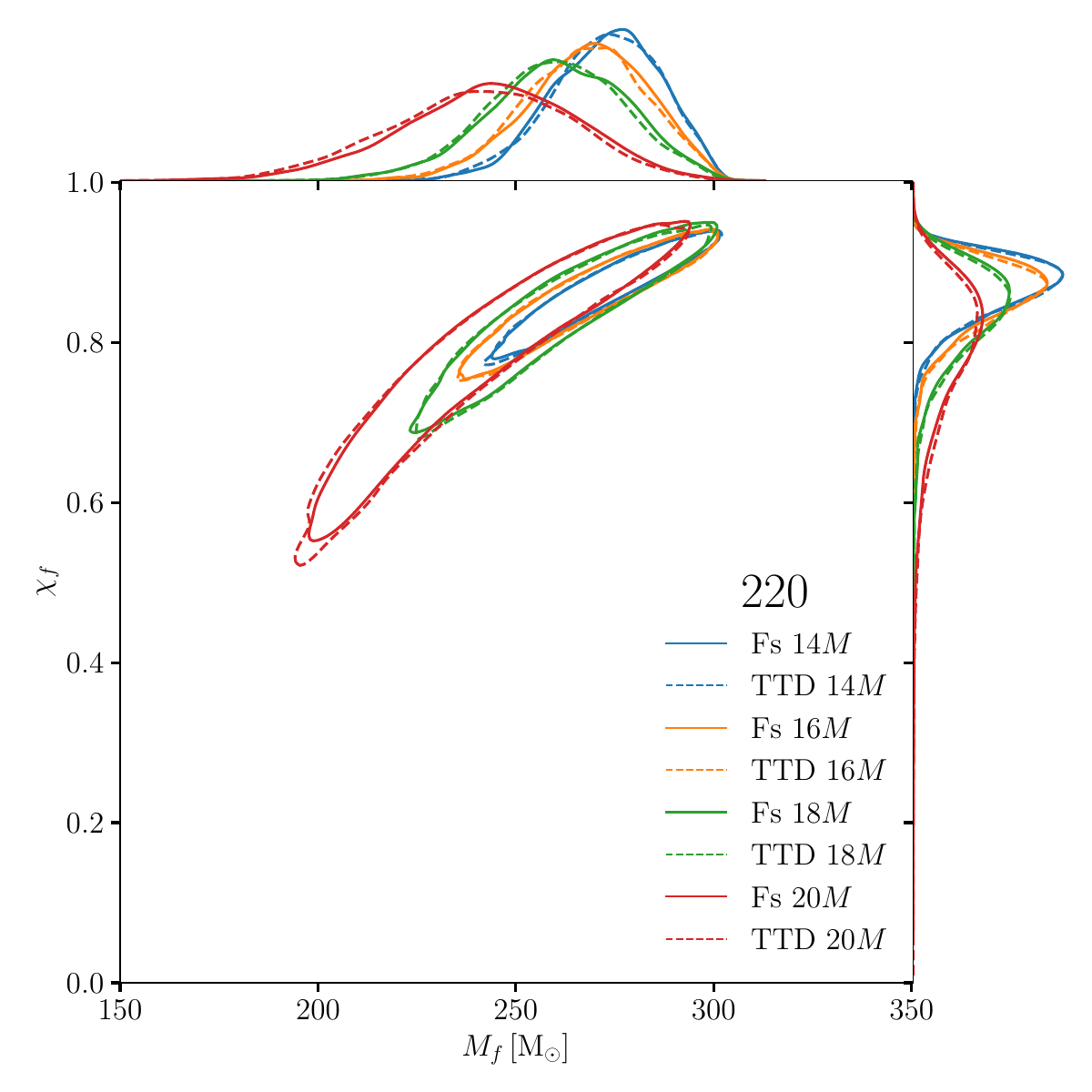}
\end{subfigure}\\
\begin{subfigure}[b]{0.45\linewidth}
\centering
\includegraphics[width=\textwidth,height=7.5cm]{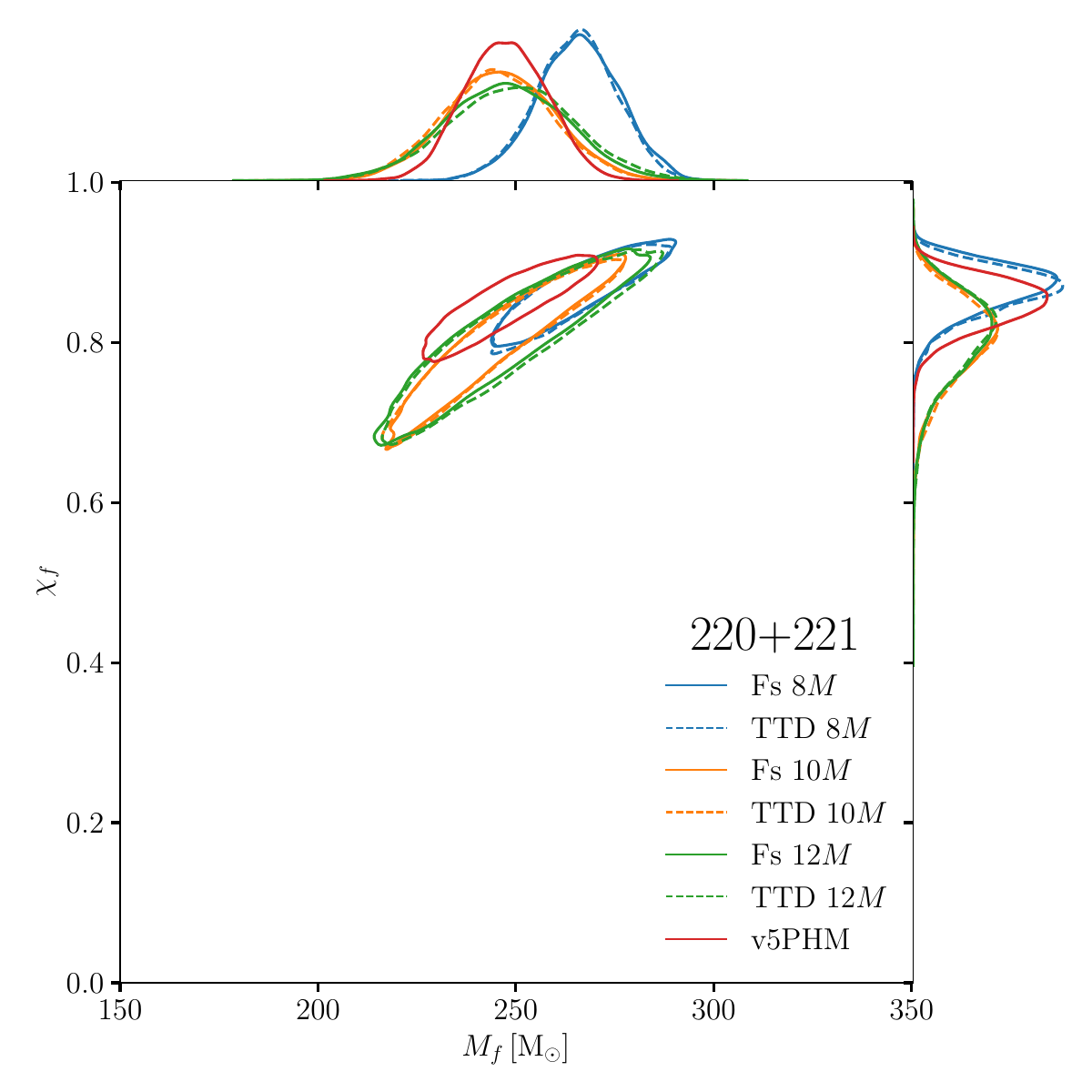}
\end{subfigure}%
\begin{subfigure}[b]{0.45\linewidth}
\centering
\includegraphics[width=\textwidth,height=7.5cm]{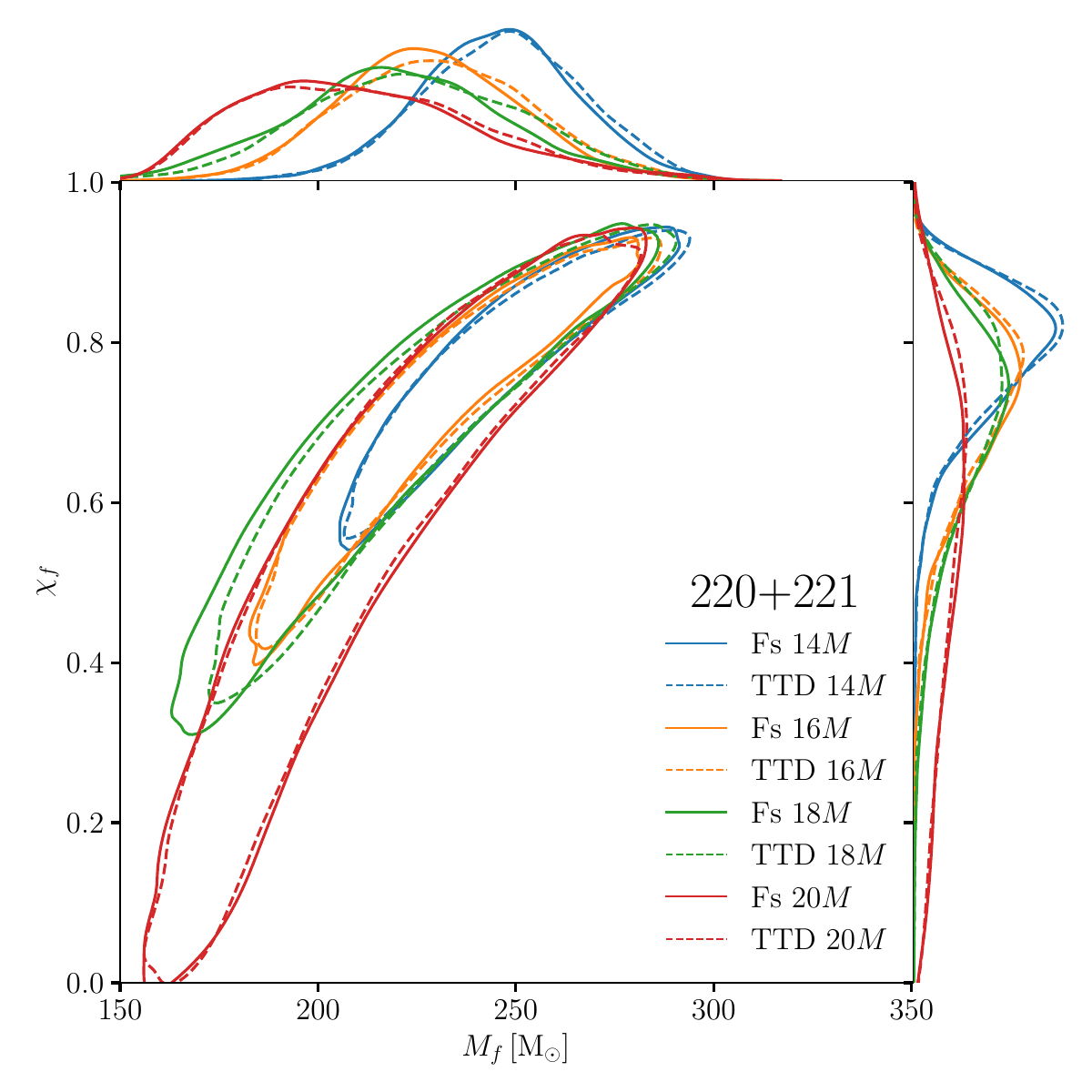}
\end{subfigure}\\
\begin{subfigure}[b]{0.45\linewidth}
\centering
\includegraphics[width=\textwidth,height=7.5cm]{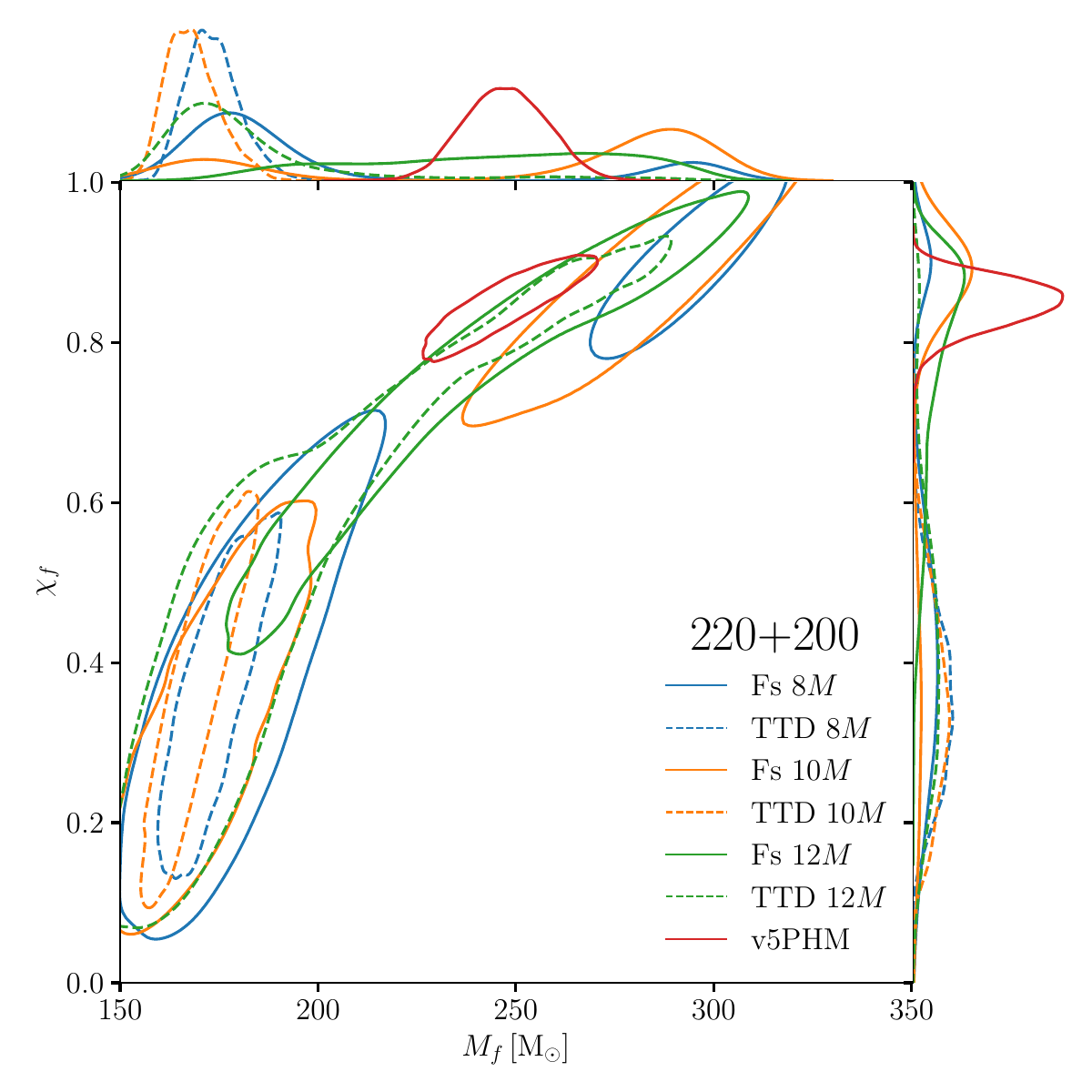}
\end{subfigure}%
\begin{subfigure}[b]{0.45\linewidth}
\centering
\includegraphics[width=\textwidth,height=7.5cm]{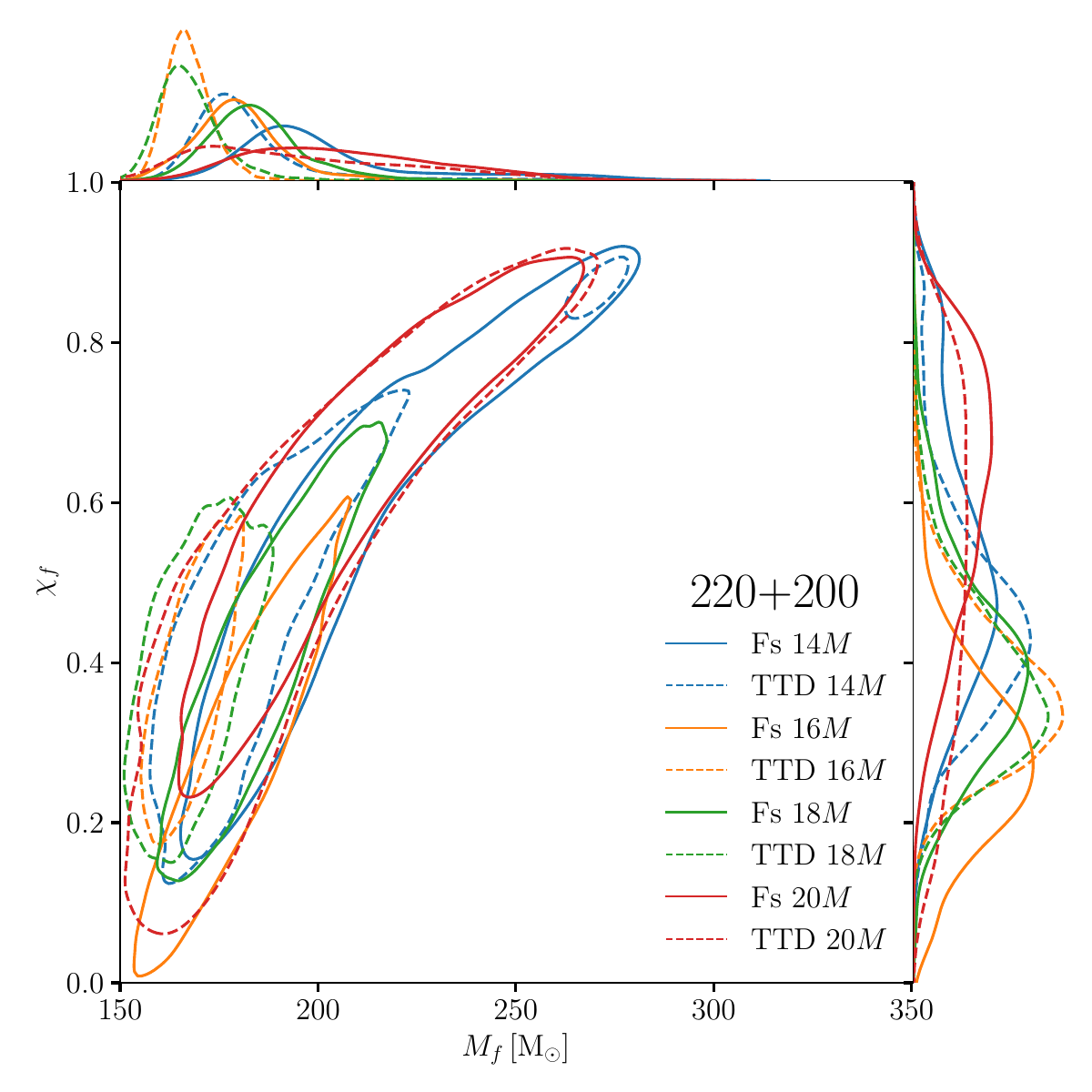}
\end{subfigure}\\
\caption{
The posterior distributions of the redshifted final mass $M_f$, and the final spin $\chi_f$, of the GW231028 remnant were obtained utilizing the $\fs$ method (soild contours) and the \ac{TTD} method (dashed contours). 
The top panels show results based on the fundamental mode only analyses at various start times, i.e., from $\Delta t=8\,M$ to $20\,M$.
The middle and bottom panels show results based on combinations $220+221$ and $220+200$, respectively.
The label `v5PHM' corresponds to the LIGO–Virgo–KAGRA Collaboration’s full \ac{IMR} analysis with the SEOBNRv5PHM waveform model.
Results are indicated by contours at the $90\%$ credible level, with the marginal posterior distributions for both $M_f$ and $\chi_f$ are shown in their respective top and right panels, respectively.
}\label{fig:fmfs1_m2}
\end{figure*}

\section{Supplemental Figures: the GW231028}\label{appen:b}

\begin{figure*}
\centering
\begin{subfigure}[b]{0.48\linewidth}
\centering
\includegraphics[width=\textwidth,height=8cm]{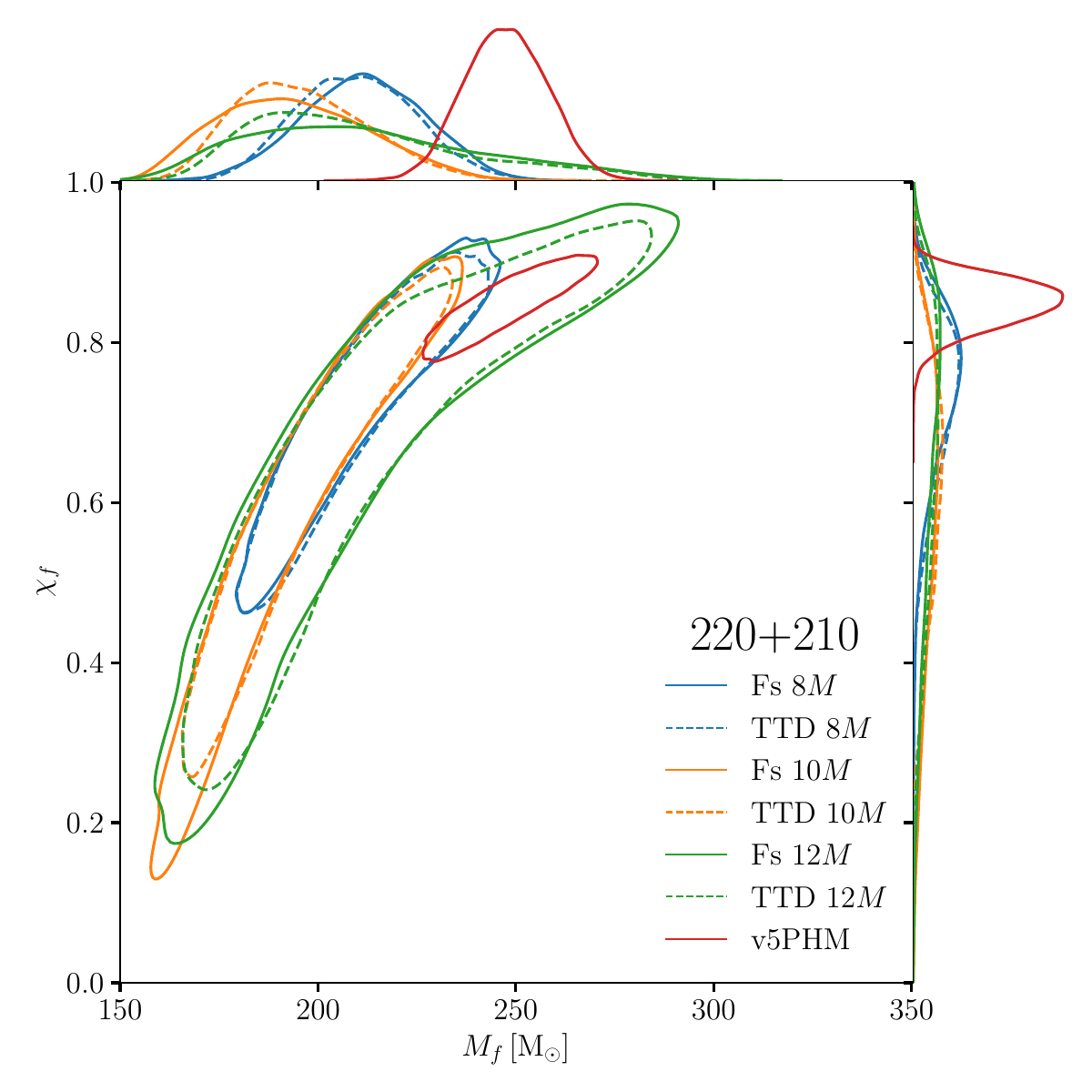}
\end{subfigure}%
\begin{subfigure}[b]{0.48\linewidth}
\centering
\includegraphics[width=\textwidth,height=8cm]{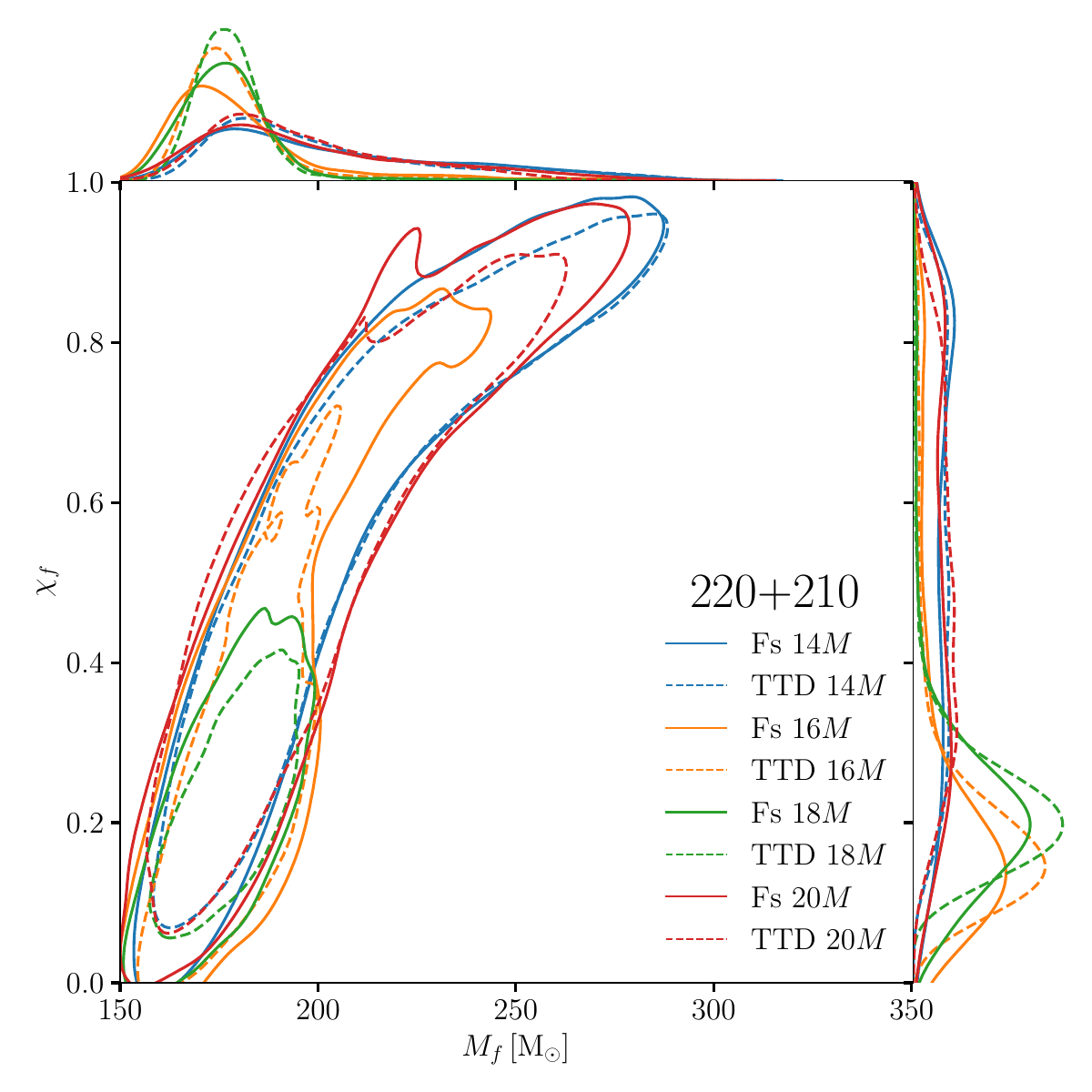}
\end{subfigure}\\
\begin{subfigure}[b]{0.48\linewidth}
\centering
\includegraphics[width=\textwidth,height=8cm]{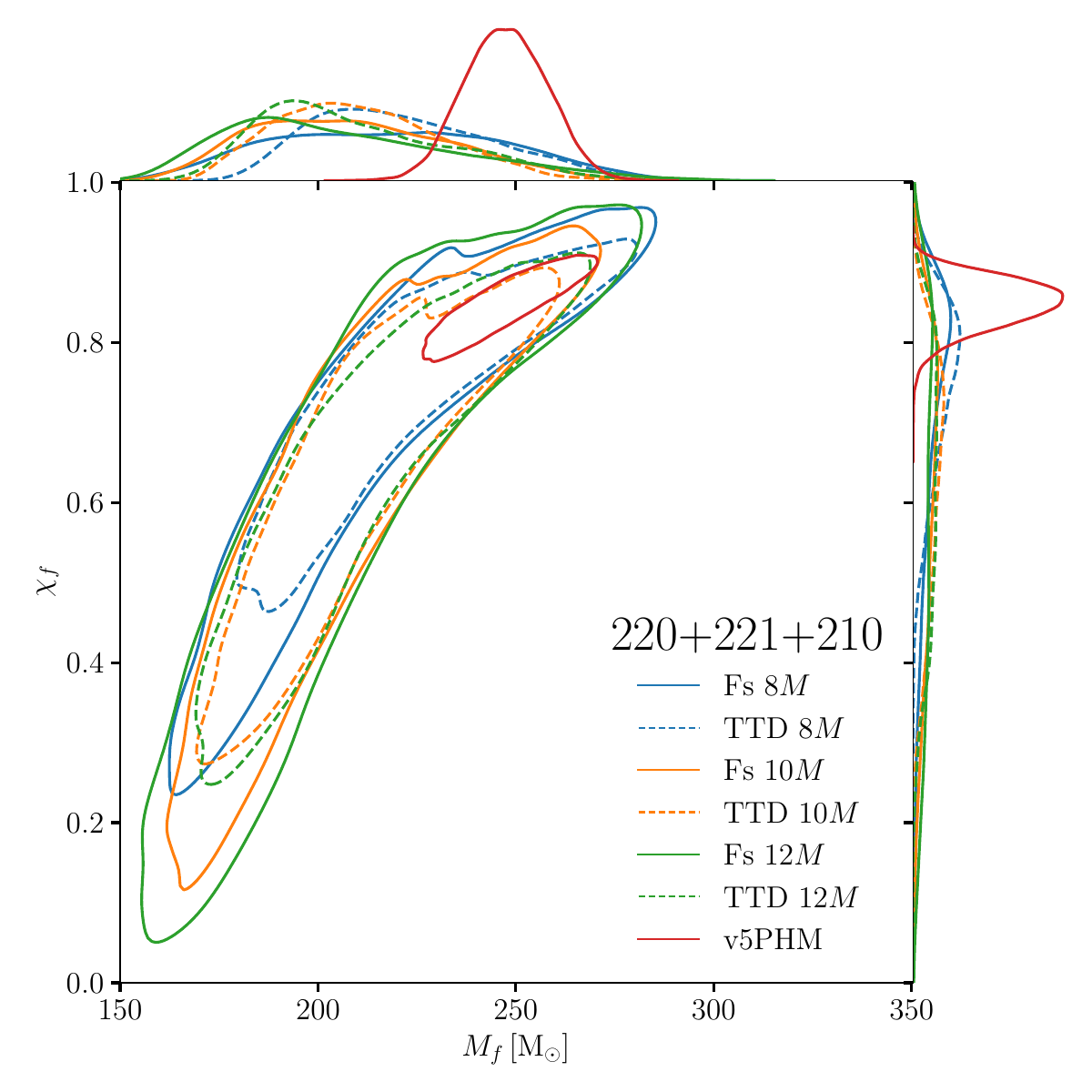}
\end{subfigure}%
\begin{subfigure}[b]{0.48\linewidth}
\centering
\includegraphics[width=\textwidth,height=8cm]{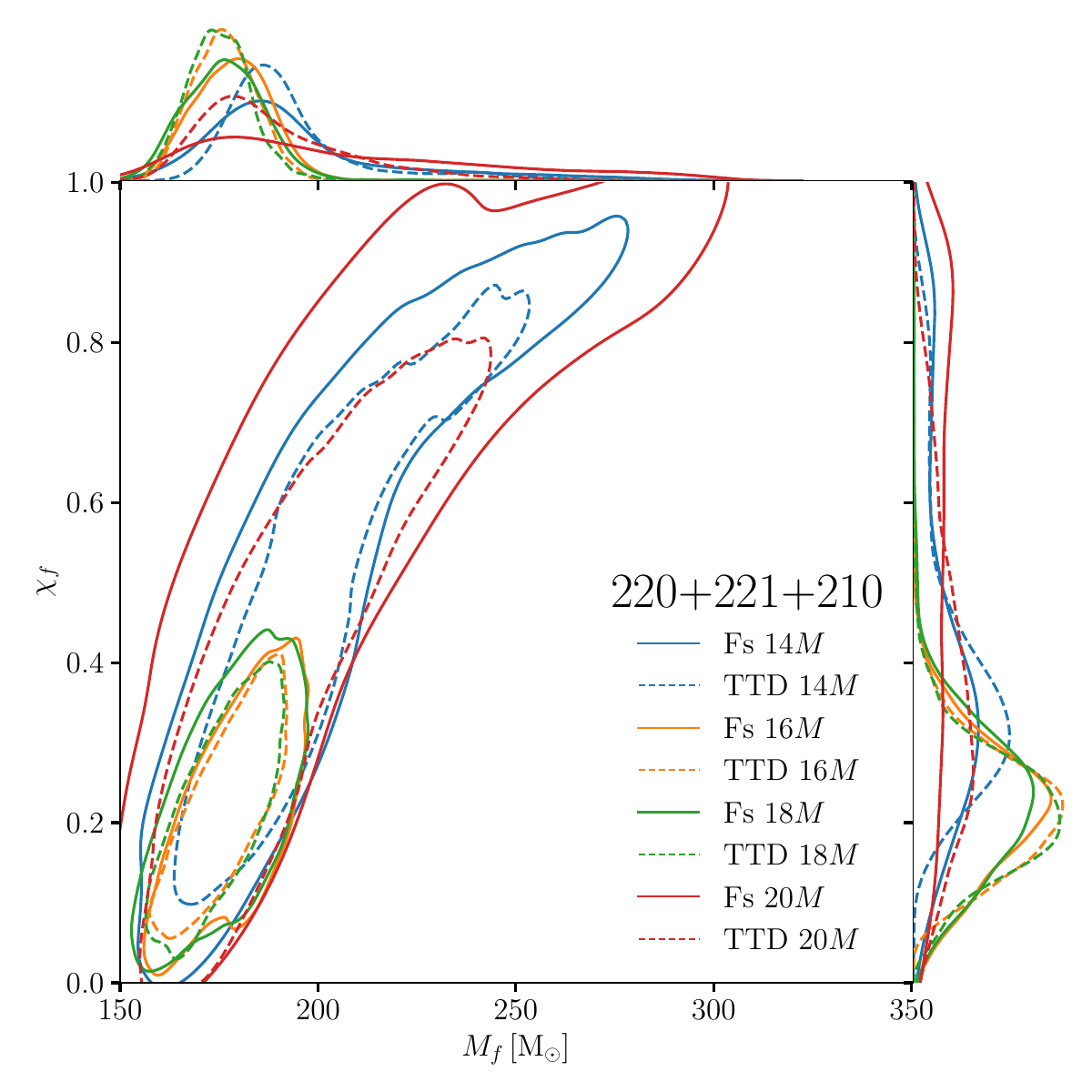}
\end{subfigure}\\
\begin{subfigure}[b]{0.48\linewidth}
\centering
\includegraphics[width=\textwidth,height=8cm]{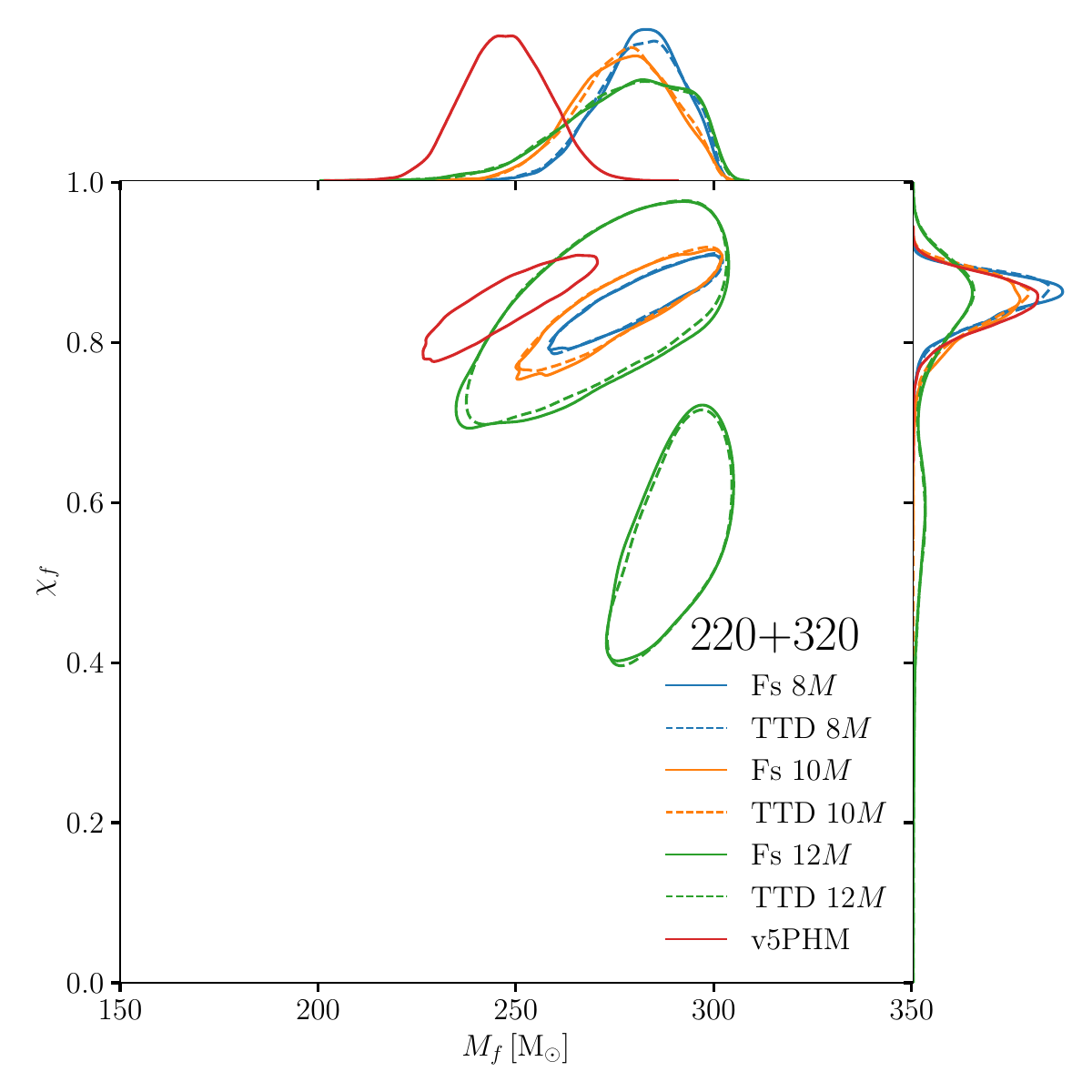}
\end{subfigure}%
\begin{subfigure}[b]{0.48\linewidth}
\centering
\includegraphics[width=\textwidth,height=8cm]{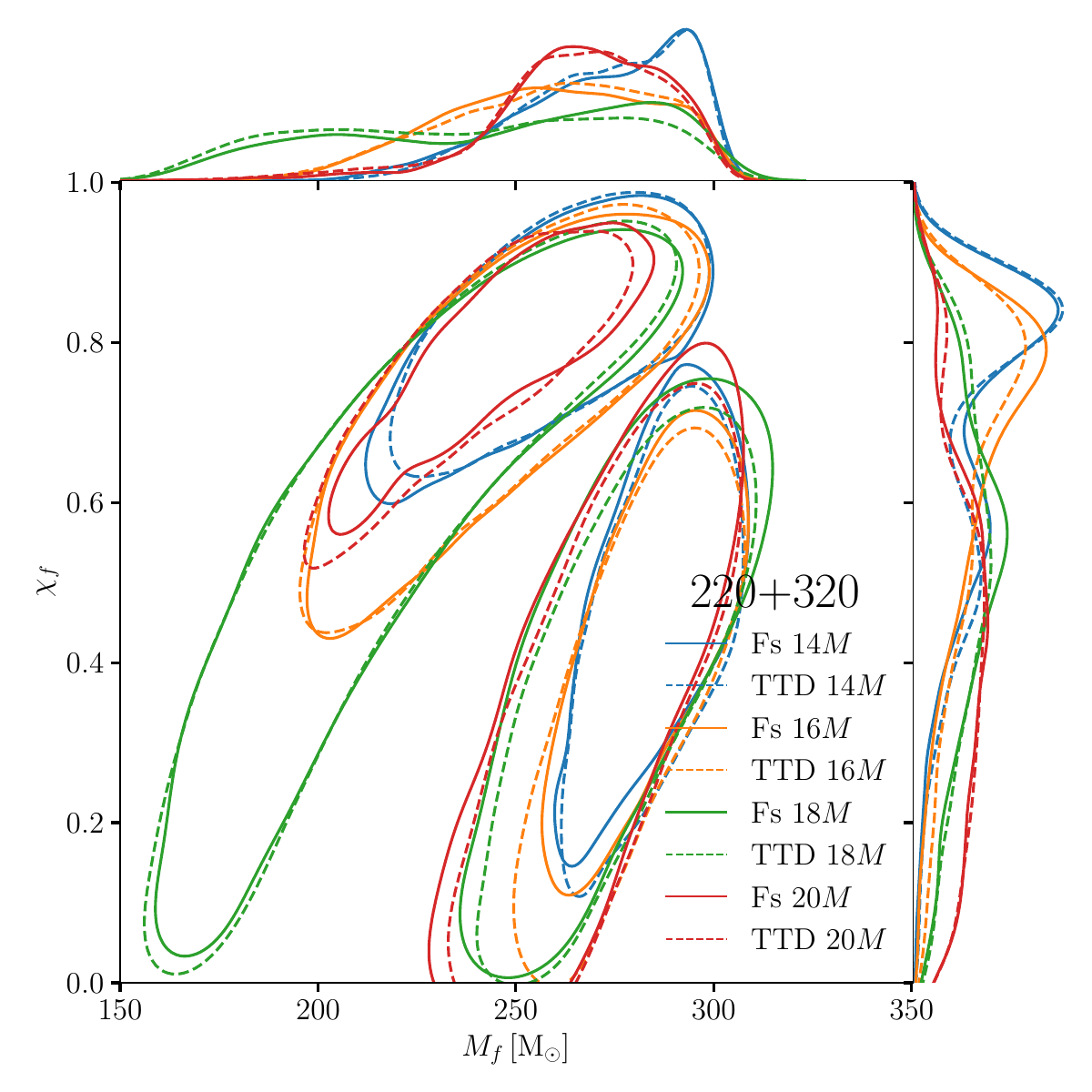}
\end{subfigure}\\
\caption{
Similar to Fig.~\ref{fig:fmfs1_m2}.
Posterior distributions of the final mass $M_f$ and final spin $\chi_f$ for the GW231028 ringdown analysis, inferred from the $220+210$, $220+221+210$, and $220+320$ mode combinations.
}\label{fig:fmfs2_m2}
\end{figure*}

\begin{figure*}
\centering
\begin{subfigure}[b]{0.48\linewidth}
\centering
\includegraphics[width=\textwidth,height=8cm]{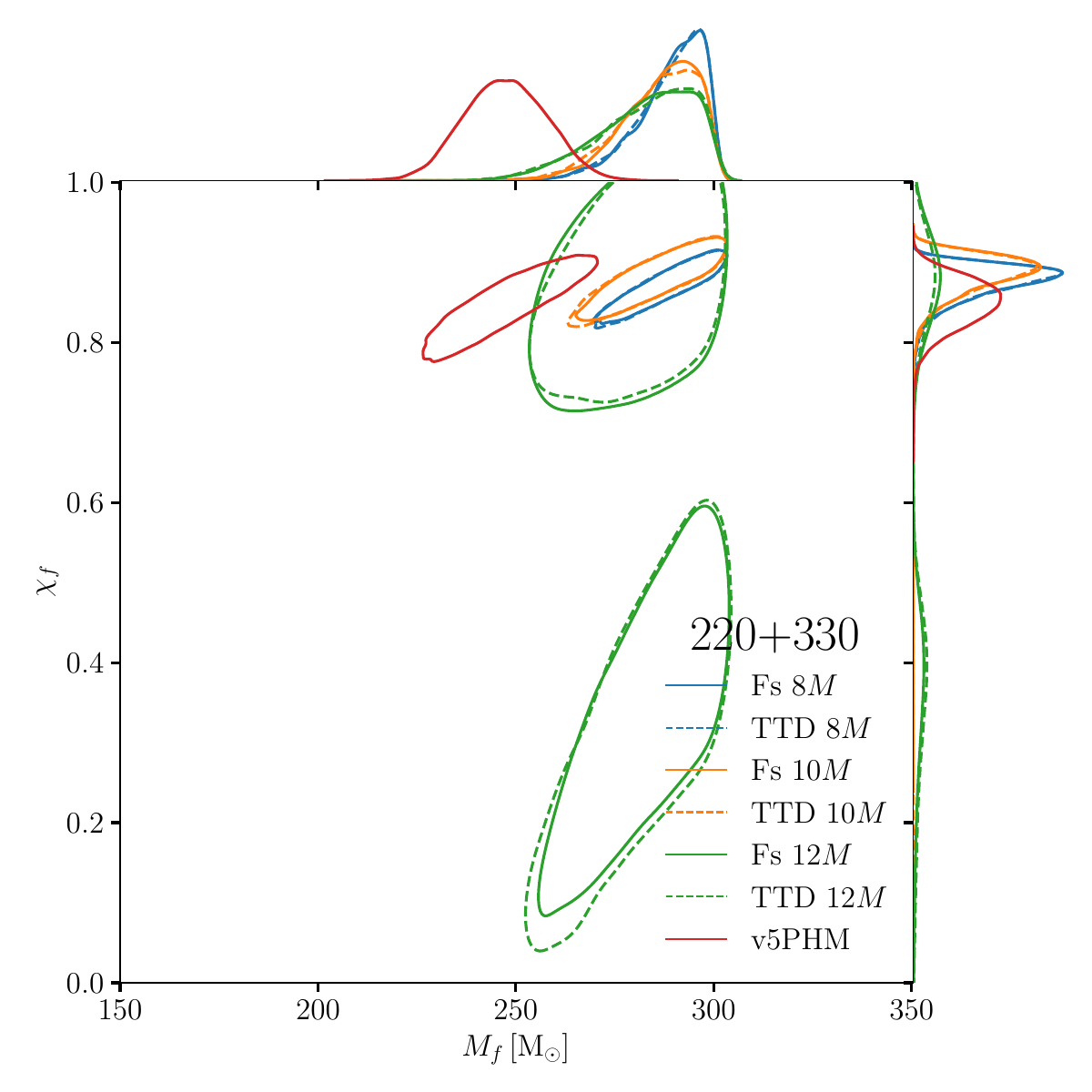}
\end{subfigure}%
\begin{subfigure}[b]{0.48\linewidth}
\centering
\includegraphics[width=\textwidth,height=8cm]{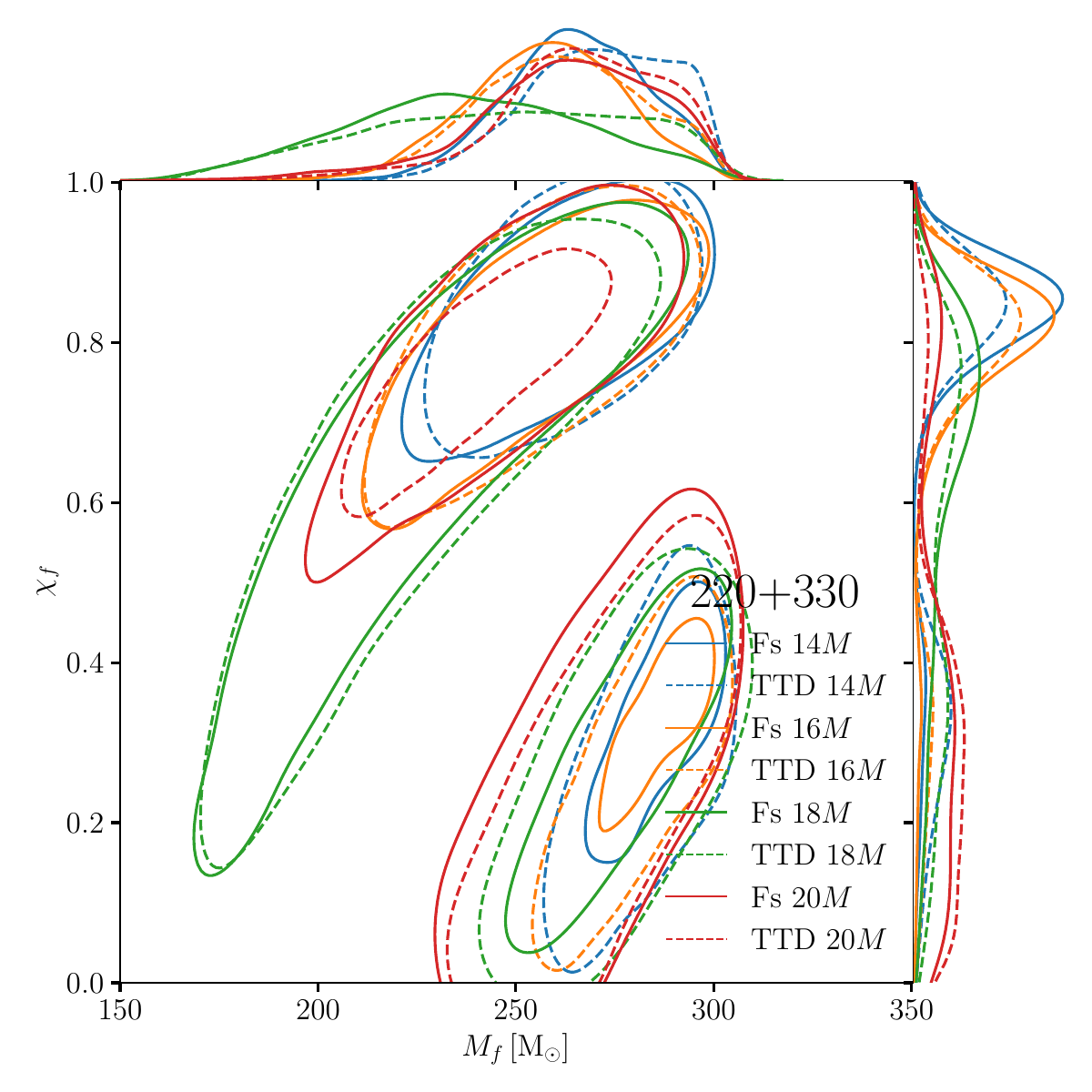}
\end{subfigure}\\
\begin{subfigure}[b]{0.48\linewidth}
\centering
\includegraphics[width=\textwidth,height=8cm]{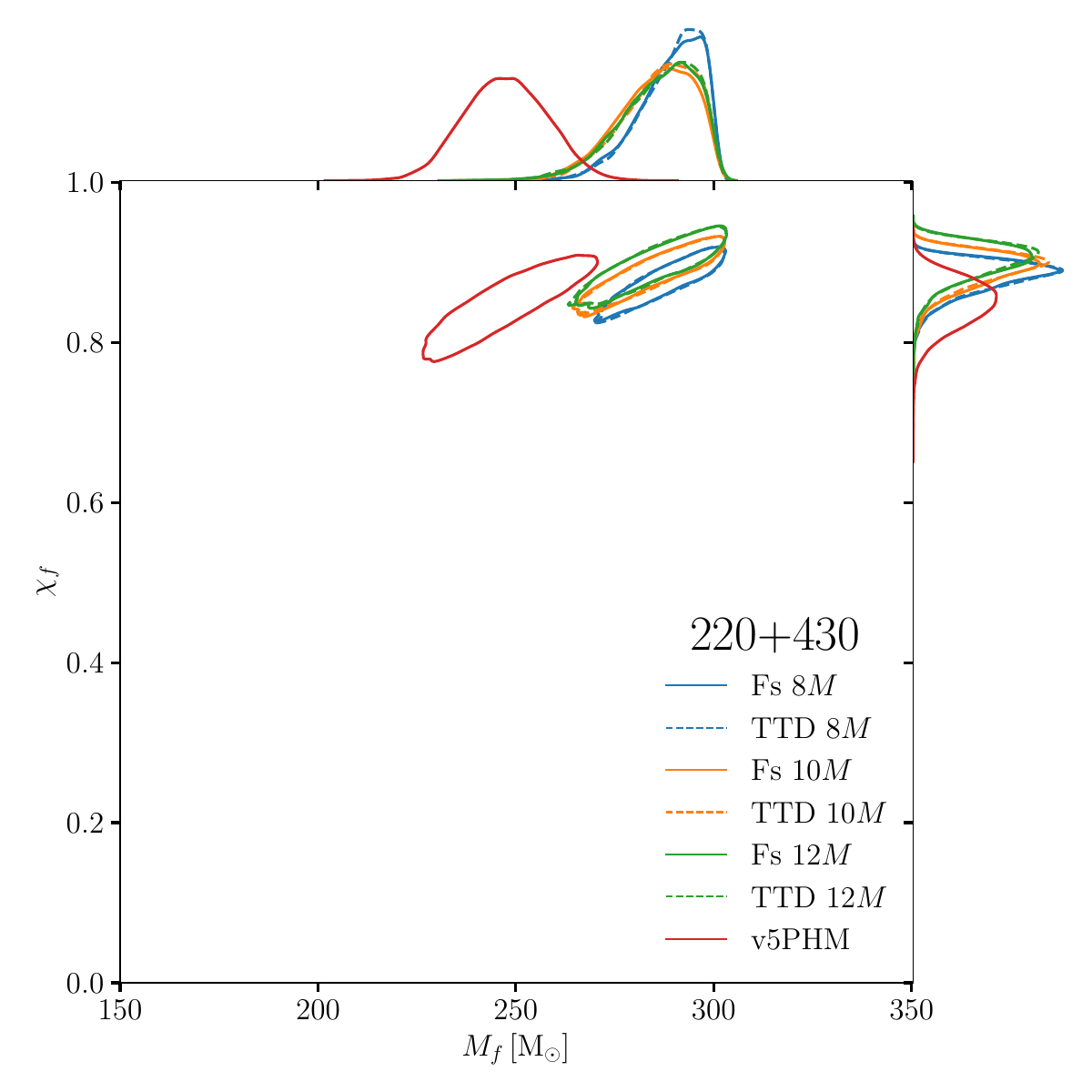}
\end{subfigure}%
\begin{subfigure}[b]{0.48\linewidth}
\centering
\includegraphics[width=\textwidth,height=8cm]{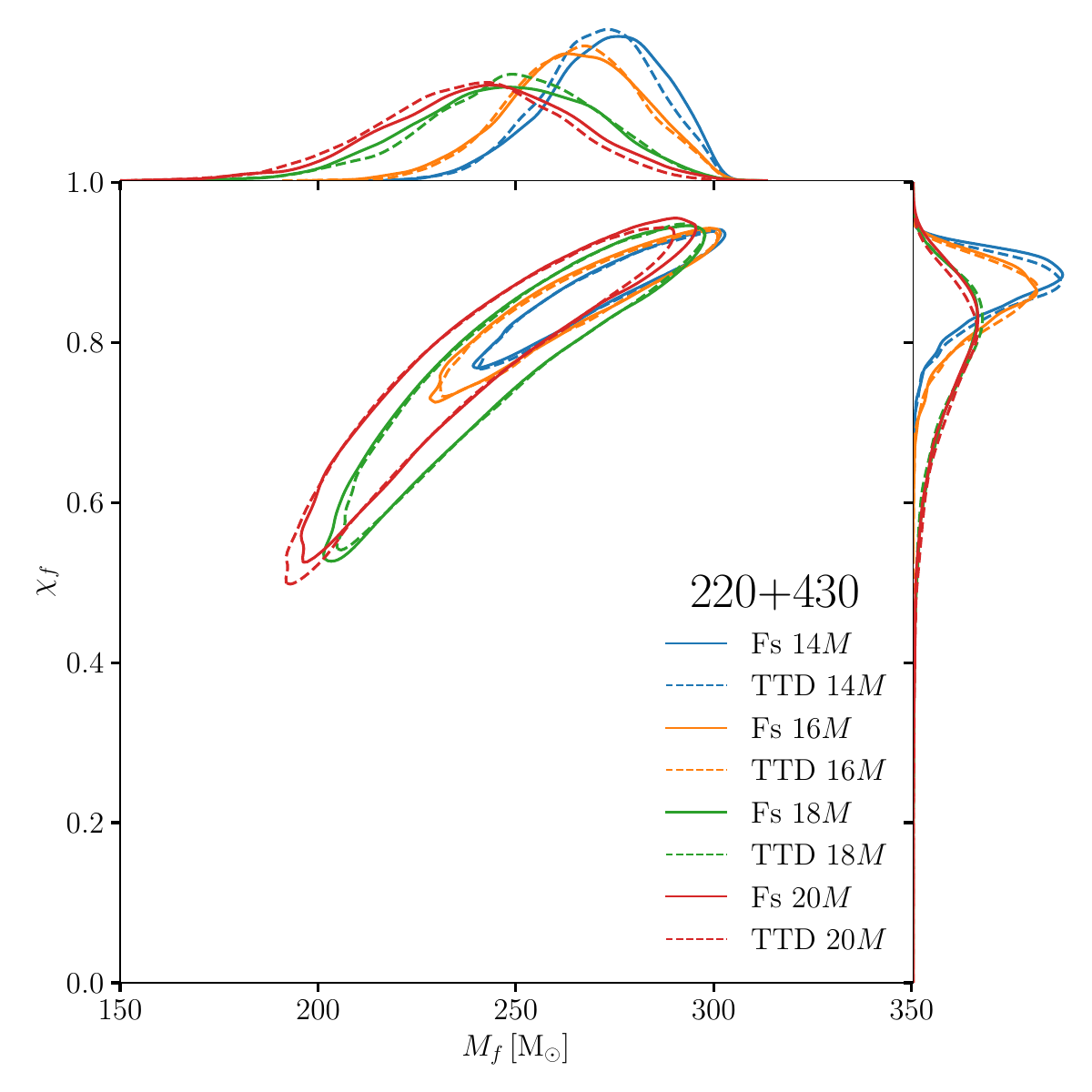}
\end{subfigure}\\
\begin{subfigure}[b]{0.48\linewidth}
\centering
\includegraphics[width=\textwidth,height=8cm]{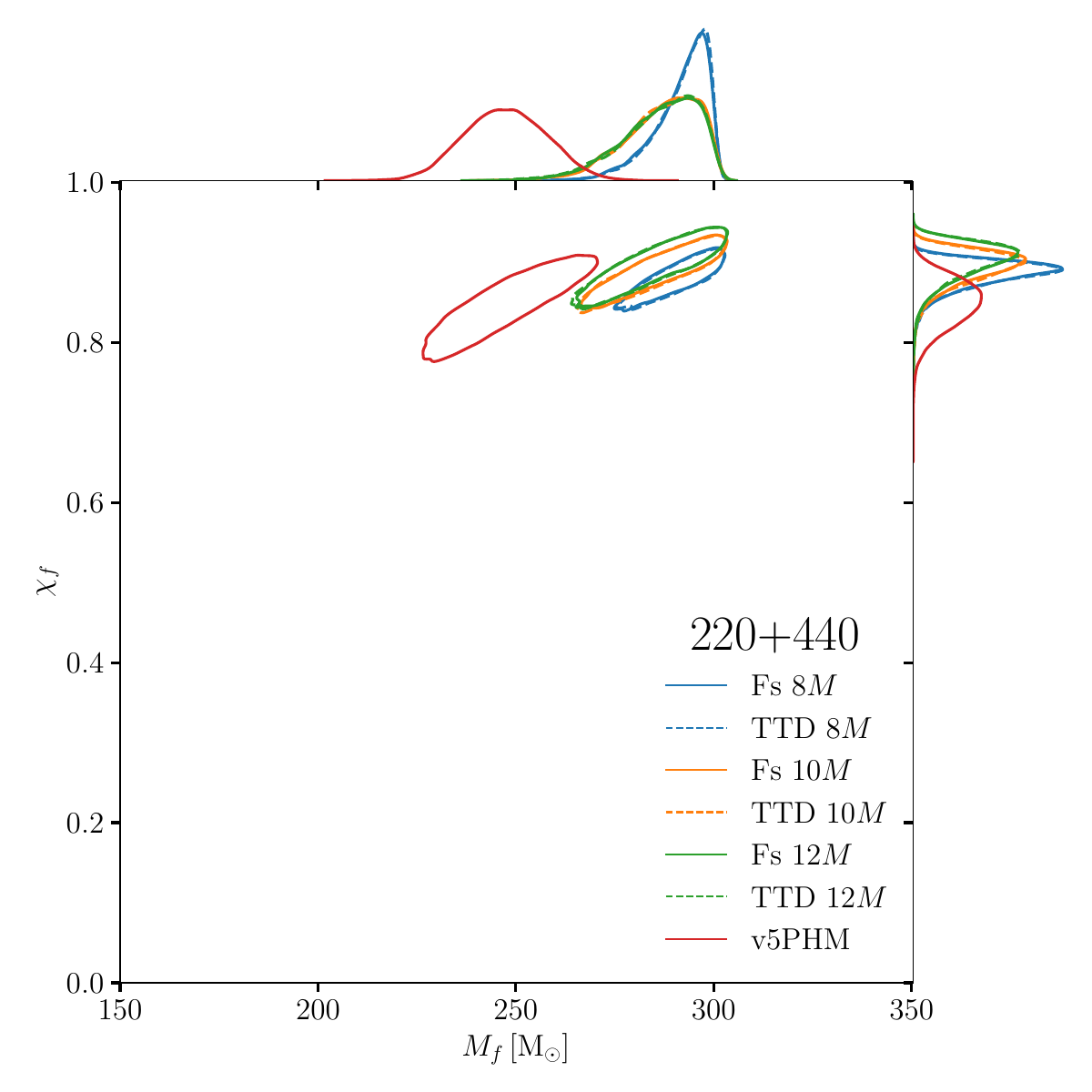}
\end{subfigure}%
\begin{subfigure}[b]{0.48\linewidth}
\centering
\includegraphics[width=\textwidth,height=8cm]{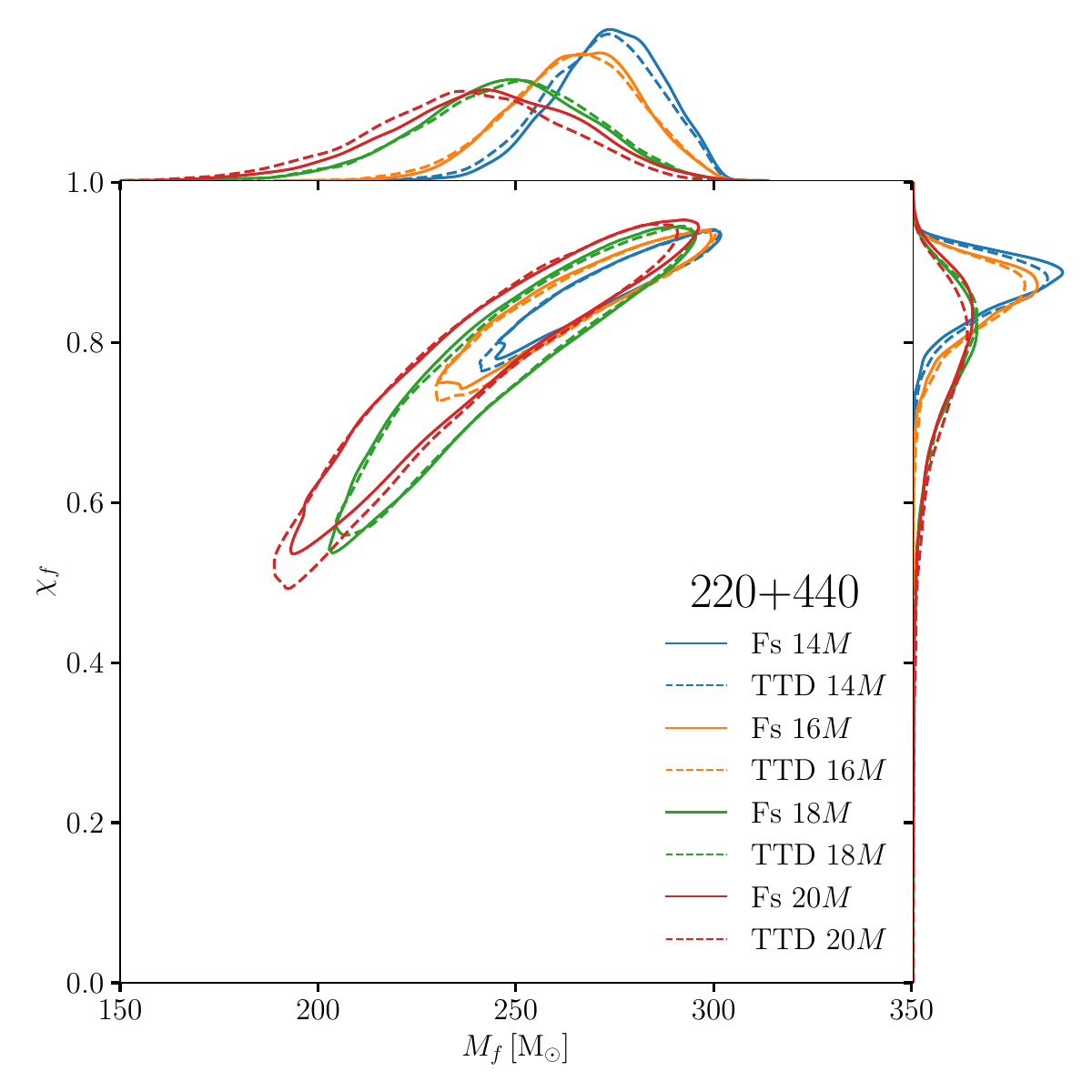}
\end{subfigure}\\
\caption{
Similar to Fig.~\ref{fig:fmfs1_m2}.
Posterior distributions of the final mass $M_f$ and final spin $\chi_f$ for the GW231028 ringdown analysis, inferred from the $220+330$, $220+430$, and $220+440$ mode combinations.
}\label{fig:fmfs3_m2}
\end{figure*}

To provide a deeper statistical perspective on the model selection results, we analyze the distributions of the log-likelihoods derived from the posterior samples. Figure~\ref{fig:lls} presents these distributions for the single-mode (220) and multi-mode (220+221, 220+210) hypotheses, comparing both the $\fs$ and \ac{TTD} methods.
Two key features are evident from this comparison. First, there is a clear separation between the likelihood distributions of the multi-mode models (orange and green) and the fundamental-only model (blue). This separation corresponds to the log-likelihood ratio ($\ln \Lambda$), providing a direct measure of the improvement in fit quality that drives the high Bayes Factors reported in the main text.
Second, the plot highlights the performance difference between the two sampling approaches. The $\fs$ method (solid lines) yields log-likelihood distributions that are noticeably narrower and shifted to slightly higher values compared to the \ac{TTD} method (dashed lines). This behavior indicates that the $\fs$ method, by analytically maximizing over amplitudes and phases, achieves tighter convergence and provides more stringent constraints on the signal parameters.
We further provide a comprehensive comparison of the posterior distributions for the remnant mass $M_f$ and spin $\chi_f$ across different ringdown models, extending the analysis discussed in the main text. The results, obtained using the \ac{TTD} and $\fs$ methods on the GW231028 ringdown signal, are displayed for all mode combinations (including the fundamental-mode-only case) as a function of the start time $\Delta t \in [8, 20] M$ in Figs.~\ref{fig:fmfs1_m2}, \ref{fig:fmfs2_m2}, and \ref{fig:fmfs3_m2}.

\begin{figure*}
\centering
\includegraphics[width=0.88\textwidth,height=12cm]{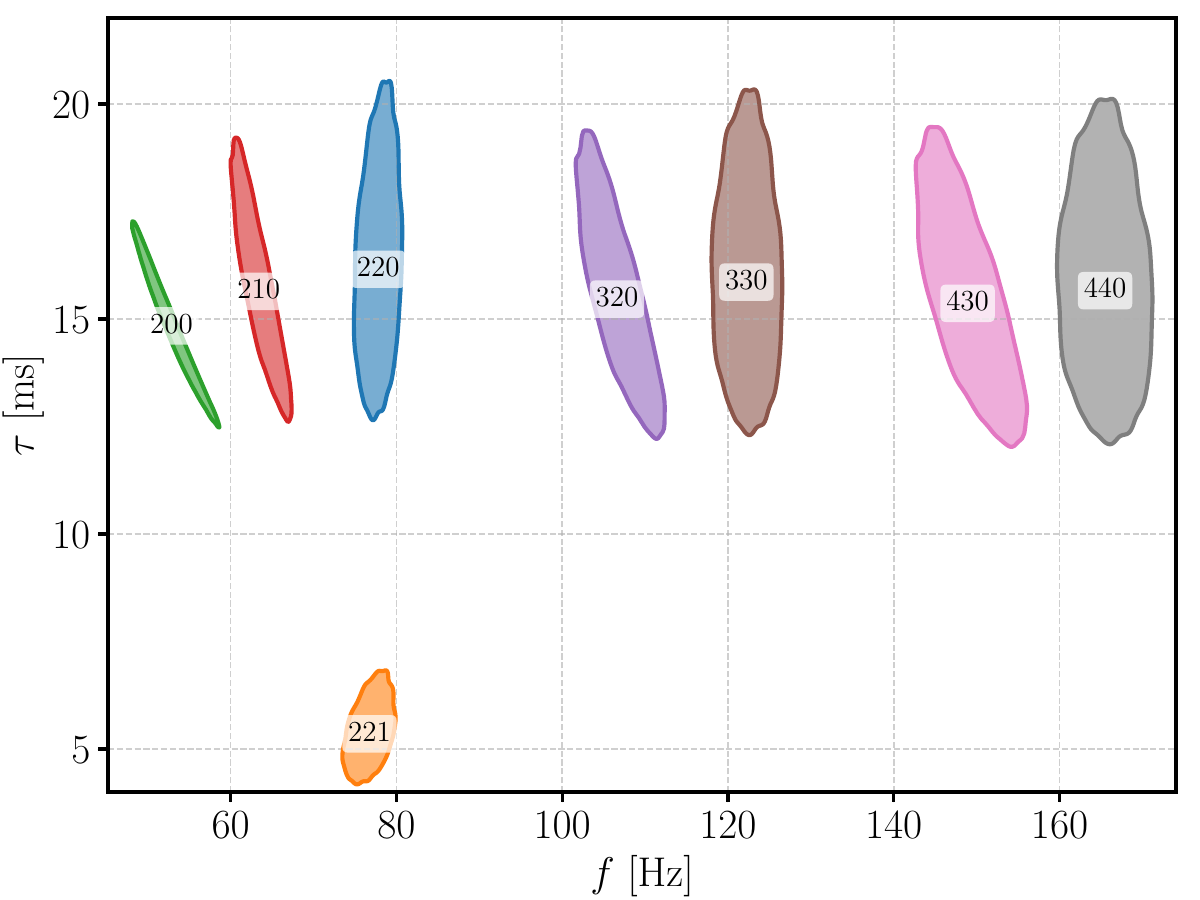}
\caption{
Predicted joint posterior distributions of the oscillation frequencies ($f$) and damping times ($\tau$) for various \acp{QNM} derived from the Kerr hypothesis. The distributions are computed using the remnant mass $M_f$ and dimensionless spin $\chi_f$ inferred from the GW231028 ringdown analysis of the $220+221$ model at a start time of $\Delta t = 10\,M$.
}\label{fig:ftaus}
\end{figure*}

Overall, excellent agreement is observed between the $\fs$ and \ac{TTD} methods across most considered mode combinations, confirming that the parameter estimation is robust against sampler systematics. However, this consistency does not extend to all mode combinations, specifically the $220+200$ and $220+320$ models, which warrant closer inspection.
As noted in the main text, the $220+200$ model exhibits discrepancies between the two methods. This is likely due to the proximity of the $(2,0)$ frequency to that of the $(2,1)$ mode (cf. Fig.~\ref{fig:ftaus}), causing the \ac{TTD} method to become trapped in a local likelihood maximum associated with the $(2,1)$ frequency, whereas the $\fs$ method is more effective at resolving closely spaced spectral peaks.

\begin{figure*}
\centering
\includegraphics[width=0.88\textwidth,height=10cm]{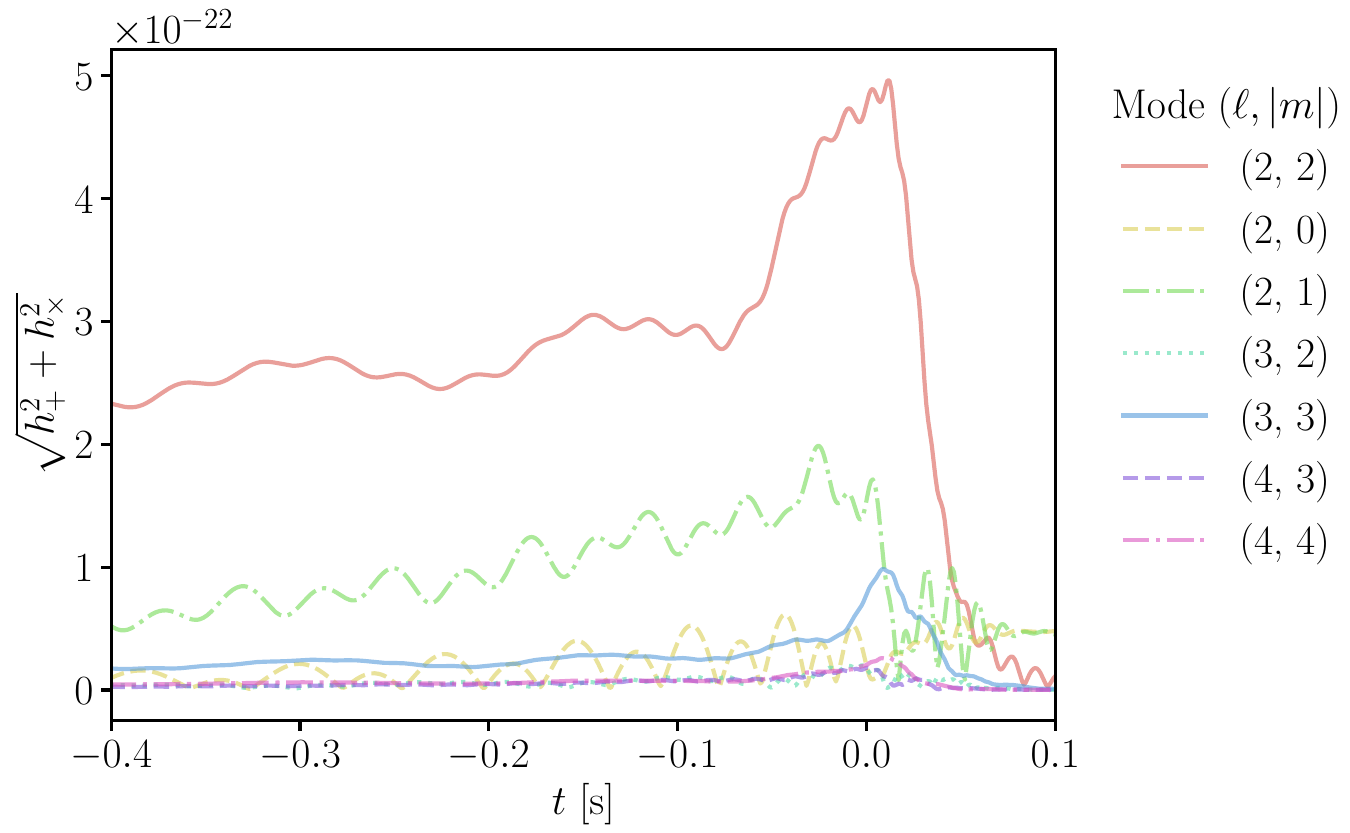}
\caption{
Time evolution of the \ac{GW} mode amplitudes for the \ac{NR} injection SXS:BBH:1282. This simulation corresponds to a \ac{BBH} system with mass ratio $q \approx 0.5$ and high spin, mimicking the source properties of GW231028. The plot displays the absolute amplitude of the dominant $(\ell, m) = (2,2)$ mode alongside various sub-dominant modes. 
}\label{fig:amps1282}
\end{figure*}

\begin{figure*}
\centering
\includegraphics[width=0.88\textwidth,height=12cm]{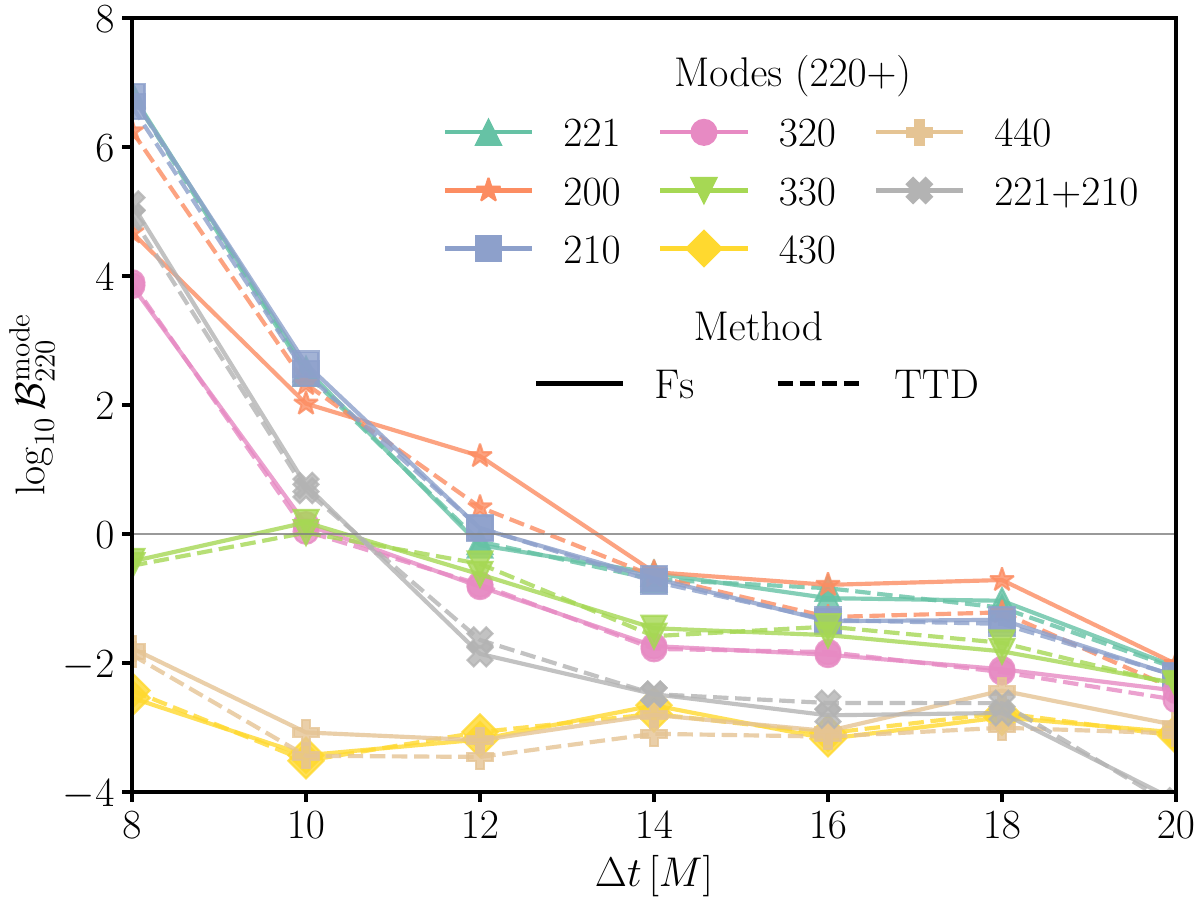}
\caption{
Bayesian model selection results for the \ac{NR} injection SXS:BBH:1282. Similar to Fig.~1 in the main text, this plot displays the $\log_{10}$ Bayes factors for various multimode hypotheses relative to the fundamental-mode-only ($220$) model across a range of start times $\Delta t$. The solid and dashed lines correspond to the results obtained using the $\fs$ and \ac{TTD} methods, respectively. Note that the injection analysis faithfully reproduces the high statistical support for the $221$ and $210$ modes observed in the real event, validating the pipeline's ability to recover these sub-dominant features in a system with similar source properties.
}\label{fig:logB1282}
\end{figure*}

Similarly, regarding the $(3,2)$ mode, although the analysis yields ``decisive" Bayes factors, we interpret this as a false positive driven by spectral leakage rather than a genuine physical detection. As illustrated in Fig.~\ref{fig:ftaus}, the oscillation frequency of the $(3,2)$ mode is spectrally very close to the dominant $(2,2)$ mode (distinct from the $(2,1)$ and $(2,0)$ frequencies). This proximity allows the sampler to pick up residual power from the dominant mode, inflating the evidence. This interpretation is strongly supported by the SXS:BBH:1282 injection analysis presented in Fig.~\ref{fig:amps1282} (and Fig.~\ref{fig:logB1282}). The injection results reproduce the high Bayes factors for both the $(2,0)$ and $(3,2)$ modes, mirroring the behavior seen in the real event, despite the fact that the actual physical amplitudes of these modes in the injection are negligible. Consequently, we attribute the high evidence for these modes to spectral confusion and do not claim their detection.

For the favored $220+221$ model, the inferred posteriors at early times ($\Delta t = 10, 12\,M$) encompass the full \ac{IMR} predictions derived from SEOBNRv5PHM. Notably, we observe that the posterior constraints for this model broaden as the start time increases up to $\Delta t = 10\,M$. This behavior is physically understood as a consequence of the spectral proximity of the fundamental mode and its first overtone. As the overtone amplitude decays, separating these two modes—which share very similar oscillation frequencies—becomes increasingly difficult, leading to stronger degeneracy and wider credible intervals for the remnant mass and spin.
In contrast, the fundamental-mode-only model ($220$) exhibits significant bias until $\Delta t \geq 16\,M$, highlighting the necessity of including the overtone. 

At $\Delta t\approx 8,10\,M$, the $220+210$ model suffers from substantial systematic bias by omitting the dominant overtone component. This dynamic changes as the signal evolves. We find that the $220+210$ model begins to agree with the \ac{IMR} predictions for start times around $\Delta t \approx 12,14\,M$ (see Fig.~\ref{fig:fmfs2_m2}). This improvement is driven by the rapid decay of the $221$ mode; as the overtone fades, the longer-lived $(2,1)$ harmonic becomes the primary sub-dominant feature.
However, for even later start times ($\Delta t \gtrsim 16\,M$), the inclusion of the $210$ mode itself begins to introduce biases. This is because the $210$ mode has also decayed to a negligible amplitude, and its oscillation frequency is very close to that of the dominant $220$ mode (cf. Fig.~\ref{fig:ftaus}). Attempting to fit a signal where the $210$ mode is not physically present with a model that includes it leads to erroneous parameter estimation due to the difficulty in distinguishing between these closely spaced frequencies.
For the three-mode analysis ($220+221+210$), it maintains consistency with \ac{IMR} results throughout the intermediate window $\Delta t \in [8, 14]\,M$, effectively bridging the early-time overtone dominance and the late-time harmonic behavior.

We also examined higher-order spherical harmonic contributions. The inclusion of $\ell=3$ modes ($220+320$ and $220+330$) is not supported by the Bayes factors. Furthermore, the resulting posteriors for these combinations are either inconsistent with the expected remnant properties or exhibit unphysical bimodal structures, suggesting that these modes are not resolvable in the current data.

Finally, regarding the $\ell=4$ modes ($220+430$ and $220+440$), we observe that while their posterior distributions at $\Delta t \gtrsim 14\,M$ nominally overlap with the \ac{IMR} predictions, this agreement does not constitute evidence for their detection. As indicated by the Bayes factors (Fig.~1 in the main text), these modes are statistically disfavored. A comparison with the single-mode ($220$) results in Fig.~\ref{fig:fmfs1_m2} reveals that the $220+430$ and $220+440$ posteriors are essentially broadened versions of the $220$-only constraints.
This behavior is physically distinct from the overtone or harmonic degeneracies discussed earlier. As illustrated in Fig.~\ref{fig:ftaus}, the oscillation frequencies of the $(4,3)$ and $(4,4)$ modes are sufficiently separated from the dominant $(2,2)$ and sub-dominant $(2,1)$ frequencies. Due to this spectral separation, the sampler does not confuse the $\ell=4$ templates with the energy of the primary modes (unlike the $220+200$ case). Consequently, their inclusion merely introduces uninformative degrees of freedom. This results in a statistical penalty (Occam's razor) and a widening of the fundamental mode's credible intervals without shifting the posterior or resolving distinct spectral features.

In the main text, we utilized a ringdown waveform model based on spherical harmonics, which assumes the time-domain symmetry relation $h_{\ell m}(t) = (-1)^\ell h^*_{\ell -m}(t)$ (where the asterisk denotes complex conjugation). This relation implies equal amplitudes for the positive and negative $m$ modes in the frame aligned with the remnant black hole's spin. To verify that this approximation holds for the precessing source GW231028b, we performed a validation analysis using a more general model.

In this validation, we replaced the spherical harmonics with spheroidal harmonics and relaxed the symmetry constraint. We treated the amplitudes and phases of the mirror modes (e.g., $(2,2)$ and $(2,-2)$) as independent free parameters. 
We tested a range of configurations using the $\fs$ method and found consistent results across all cases. However, for clarity and relevance, we focus here on presenting these three representative scenarios:
\begin{enumerate}
\item Fundamental mode only: $(2, \pm 2, 0)$ starting at $t_0 = 20M$.
\item Fundamental + Overtone: $(2, \pm 2, 0) + (2, \pm 2, 1)$ starting at $t_0 = 10M$.
\item Fundamental + Higher Harmonic: $(2, \pm 2, 0) + (2, \pm 1, 0)$ starting at $t_0 = 8M$.
\end{enumerate}

\begin{figure*}
\centering
\begin{subfigure}[b]{0.48\linewidth}
\centering
\includegraphics[width=\textwidth,height=7cm]{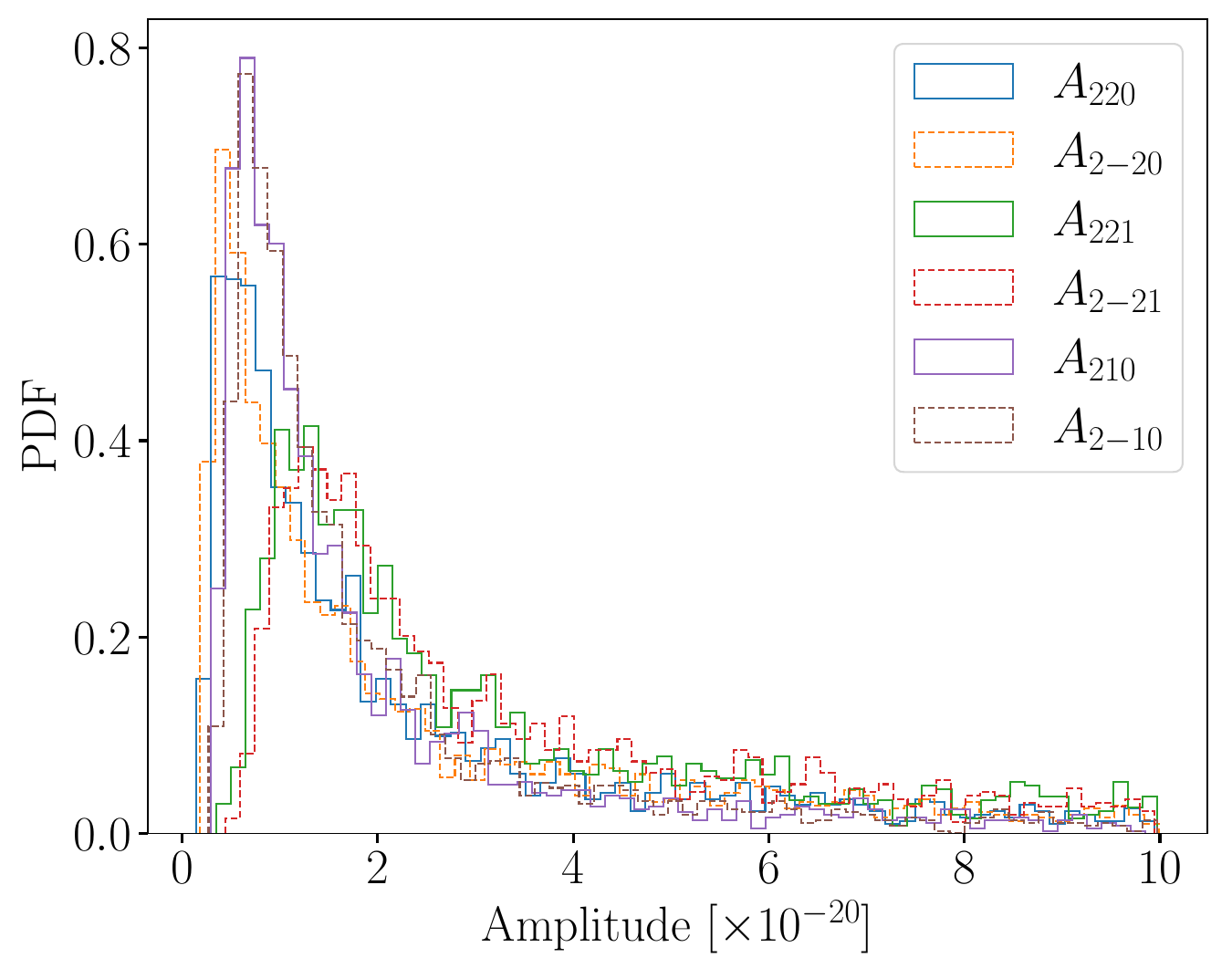}
\end{subfigure}%
\begin{subfigure}[b]{0.48\linewidth}
\centering
\includegraphics[width=\textwidth,height=8cm]{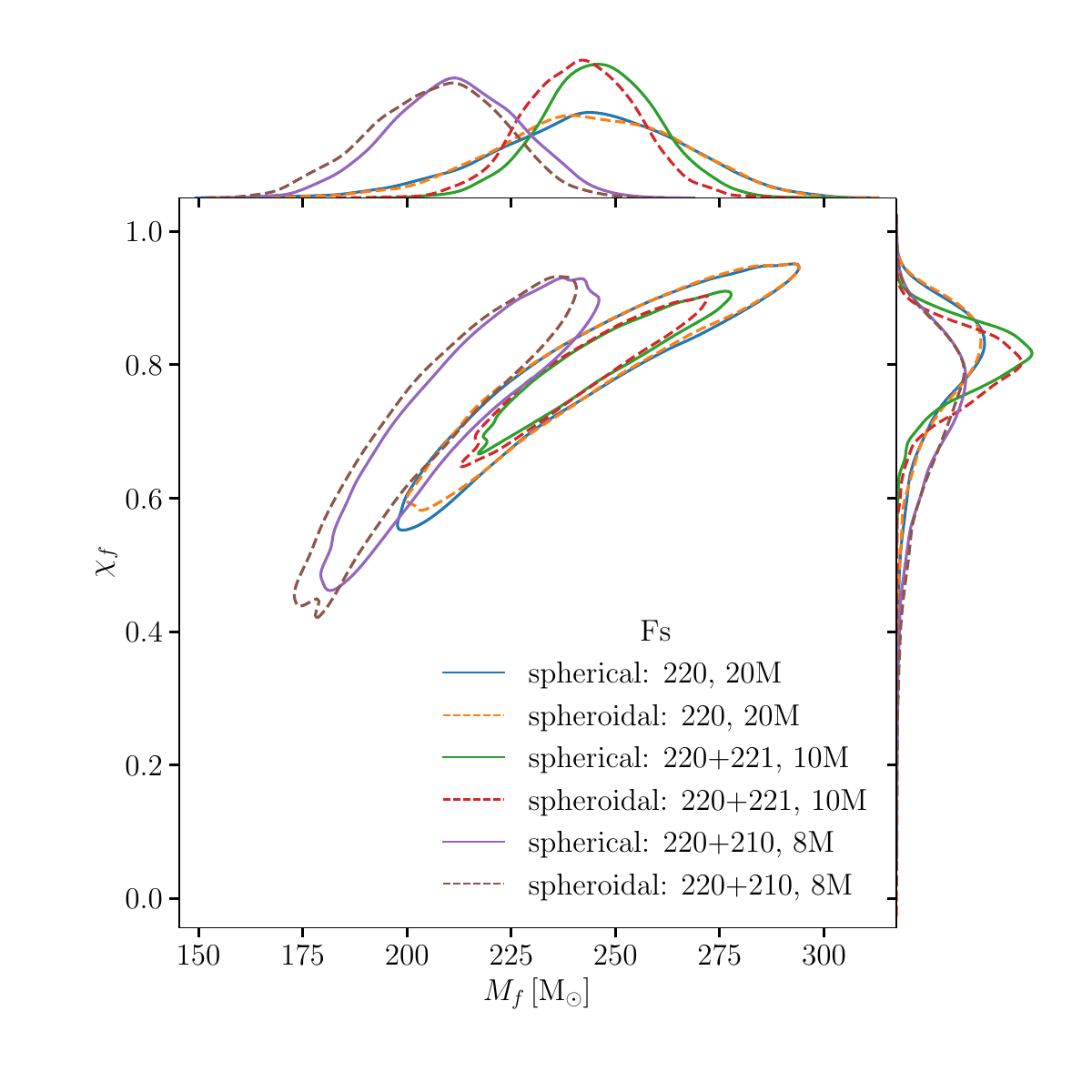}
\end{subfigure}\\
\caption{
Comparison of ringdown constraints using spherical vs. spheroidal harmonic approximations.
The left panel displays the posterior probability density functions (PDFs) for the amplitudes of the dominant modes ($A_{220}$, $A_{221}$, $A_{210}$) and their corresponding mirror modes ($A_{2-20}$, $A_{2-21}$, $A_{2-10}$) derived from the spheroidal analysis where conjugate symmetry is not assumed. The amplitude distributions for $A_{220}$ and $A_{2-20}$ are drawn from the $(2,|2|,0)+(2,|1|,0)$ analysis at $t_0=10M$; however, analogous comparisons using results from the other analyses yield consistent conclusions. The right panel compares the $90\%$ credible regions for the remnant mass $M_f$ and dimensionless spin $\chi_f$ obtained using the standard spherical harmonic model (assuming $h_{\ell m} = h_{\ell -m}^*$, solid lines) versus the fully independent spheroidal harmonic model (dashed lines). The comparisons are shown for three mode configurations: $(2,|2|,0)$ starting at $20M$, $(2,|2|,0)+(2,|2|,1)$ at $10M$, and $(2,|2|,0)+(2,|1|,0)$ at $8M$. 
}\label{fig:ampfmfs_m2}
\end{figure*}

The results are presented in Fig.~\ref{fig:ampfmfs_m2}. The right panel demonstrates that the posterior contours for the remnant mass ($M_f$) and dimensionless spin ($\chi_f$) are remarkably consistent between the ``spherical'' (symmetric) and ``spheroidal'' (independent) models. We quantified this agreement using the Jensen-Shannon Divergence (JSD) and found JSD values $< 0.03$ for remnant parameters, indicating negligible information loss or bias introduced by the symmetry assumption.
The left panel of Fig.~\ref{fig:ampfmfs_m2} explicitly compares the recovered amplitudes of the positive and negative $m$ modes ($A_{220}$ and $A_{2-20}$ are chosen from the $10M$ analysis; however, analogous comparisons using results from the other analyses yield consistent conclusions.). The distributions overlap significantly, suggesting that at the \ac{SNR} of this event ($\rho \approx 17$), any potential asymmetry in the excitation of mirror modes caused by precession is not statistically resolvable. Consequently, the use of the standard symmetric waveform model in the main text is robust.

\section{Validation with Numerical Relativity Injection}\label{appen:c}
\label{sec:injection}

To strictly validate the robustness of our multimode detections and the reliability of our analysis pipeline, we performed a targeted injection study using a \ac{NR} waveform.

\subsection{Injection Parameters}
We selected the waveform \texttt{SXS:BBH:1282} from the SXS catalog. This simulation was chosen because its physical parameters closely mimic the source properties inferred for GW231028, particularly the high remnant spin and mass ratio. The system models a binary black hole merger with a mass ratio $q \approx 2$ and large dimensionless component spins of $\chi_1 \approx 0.88$ and $\chi_2 \approx 0.88$. At the reference frequency, the binary exhibits an effective inspiral spin of $\chi_{\mathrm{eff}} \approx 0.45$ and a precession spin parameter of $\chi_p \approx 0.74$, indicating significant spin–orbit coupling. The orbital eccentricity is negligible ($e \approx 5.6 \times 10^{-4}$). The simulation covers approximately 23 inspiral orbits prior to merger.

\begin{figure*}
\centering
\begin{subfigure}[b]{0.45\linewidth}
\centering
\includegraphics[width=\textwidth,height=7.5cm]{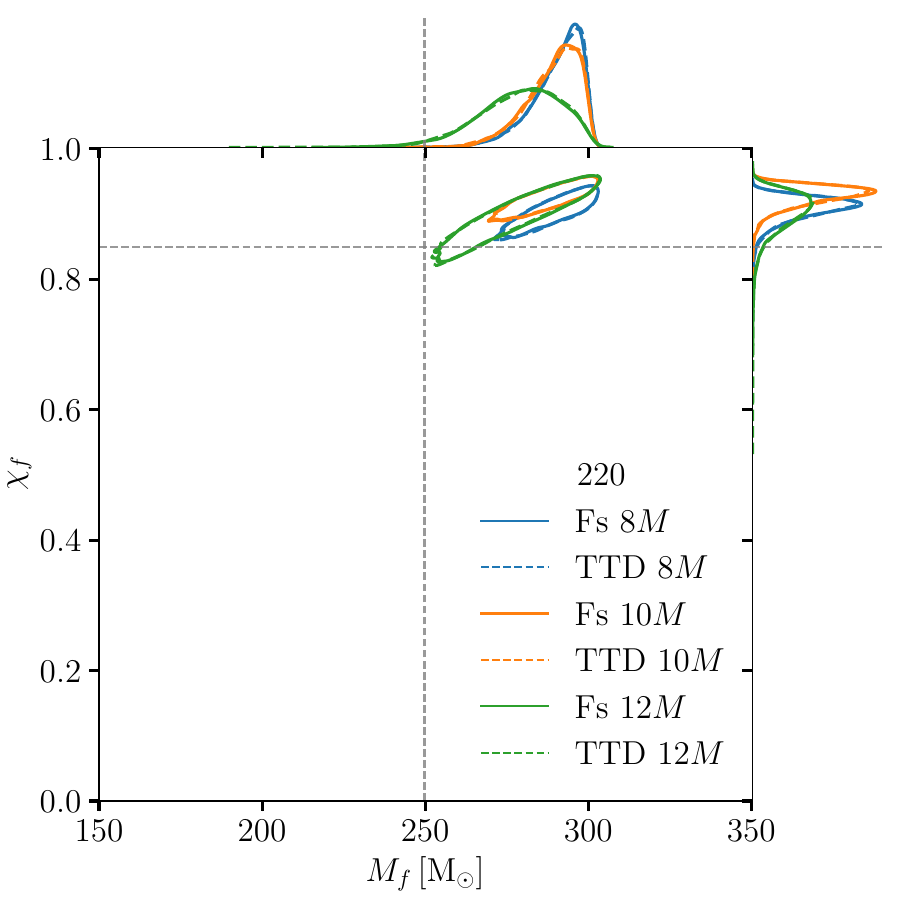}
\end{subfigure}%
\begin{subfigure}[b]{0.45\linewidth}
\centering
\includegraphics[width=\textwidth,height=7.5cm]{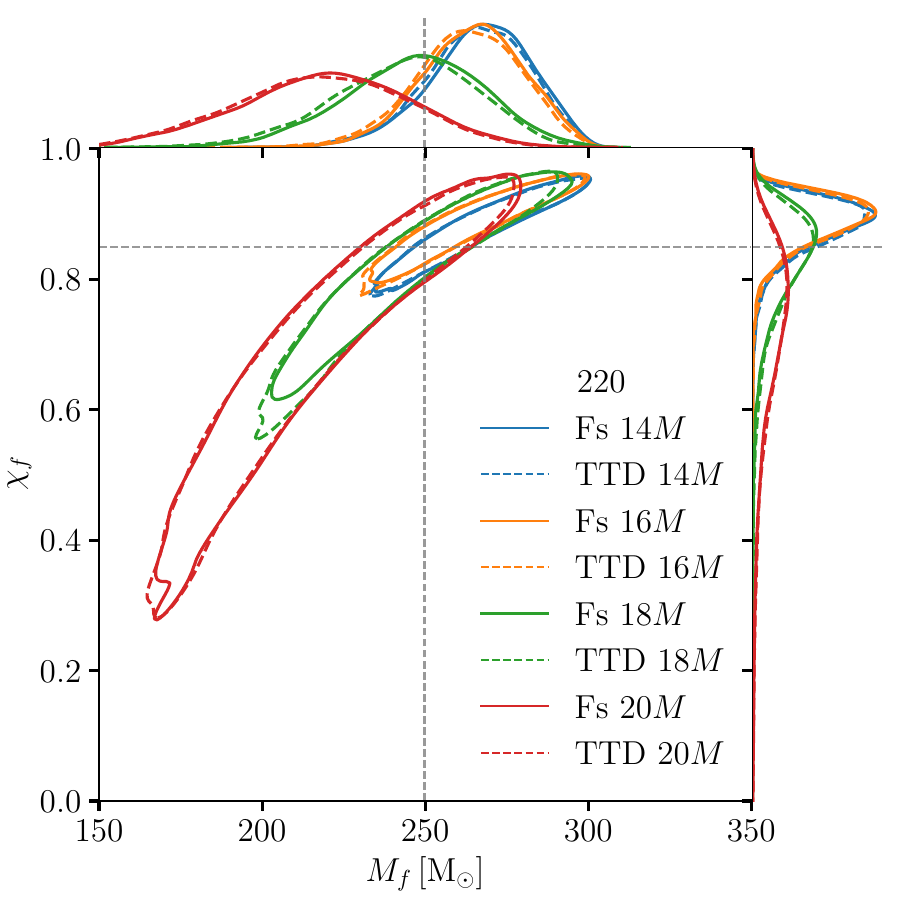}
\end{subfigure}\\
\begin{subfigure}[b]{0.45\linewidth}
\centering
\includegraphics[width=\textwidth,height=7.5cm]{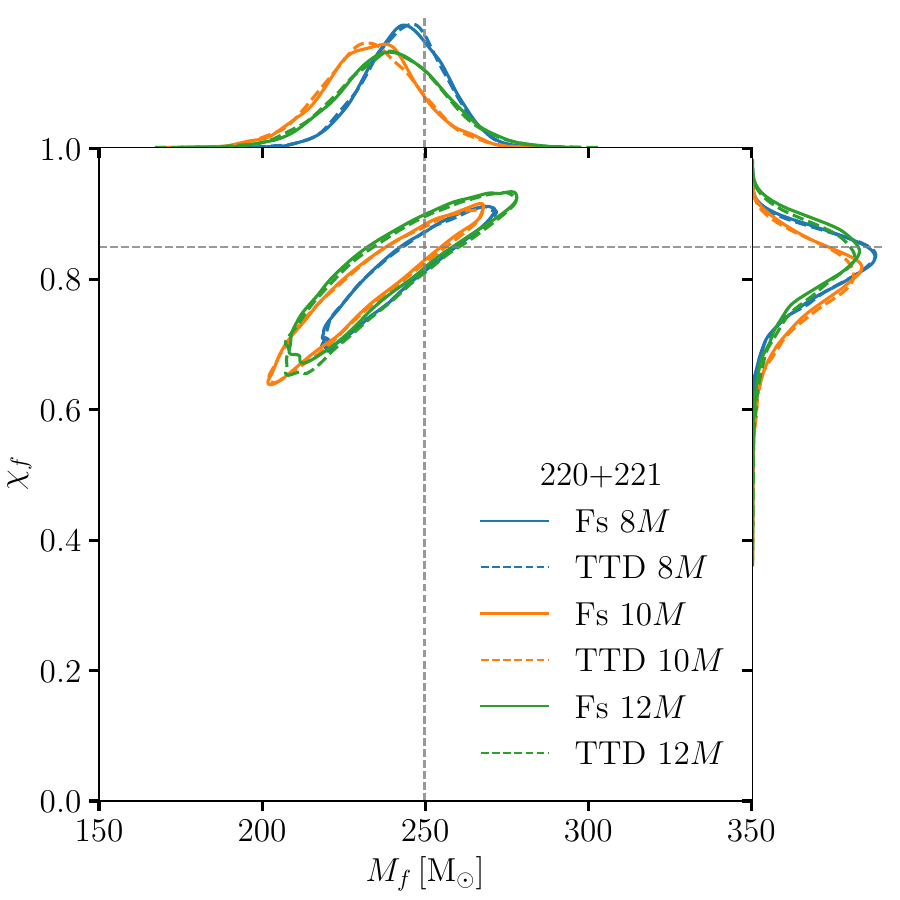}
\end{subfigure}%
\begin{subfigure}[b]{0.45\linewidth}
\centering
\includegraphics[width=\textwidth,height=7.5cm]{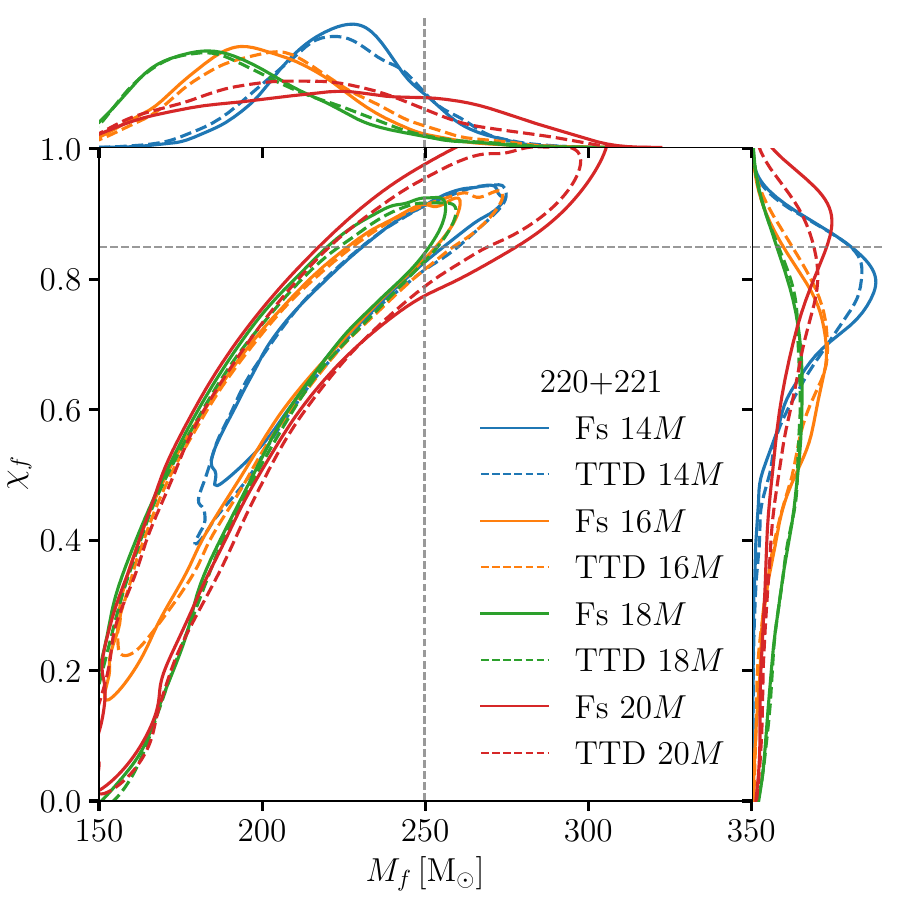}
\end{subfigure}\\
\begin{subfigure}[b]{0.45\linewidth}
\centering
\includegraphics[width=\textwidth,height=7.5cm]{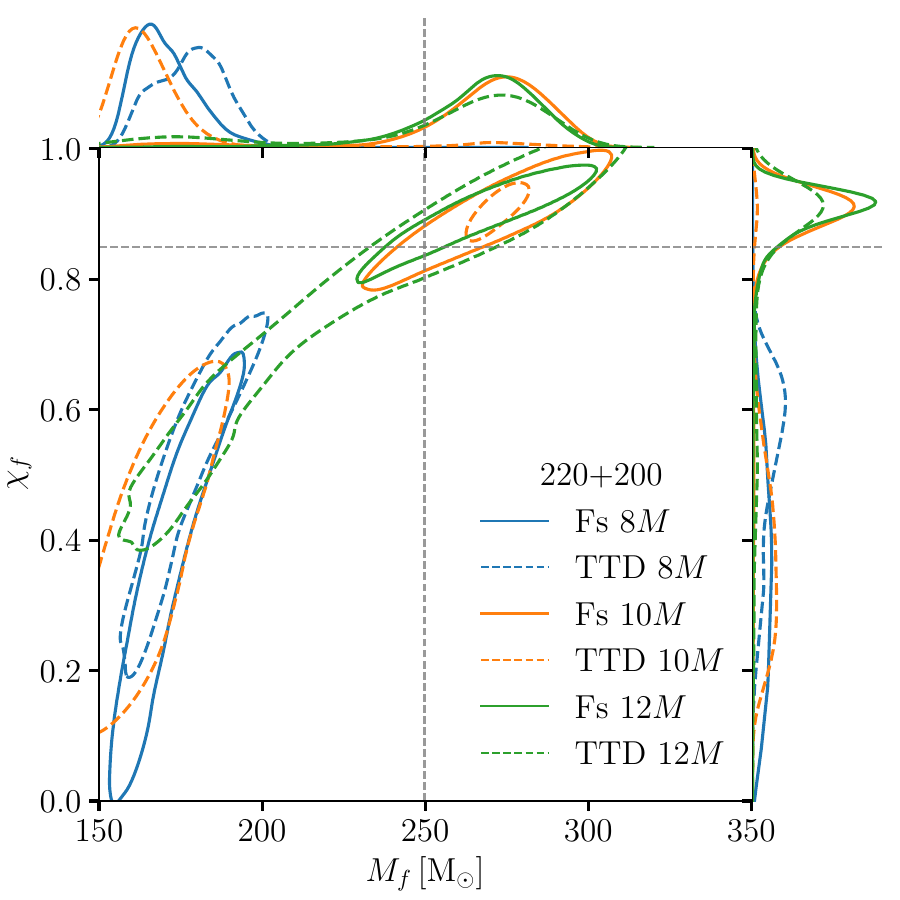}
\end{subfigure}%
\begin{subfigure}[b]{0.45\linewidth}
\centering
\includegraphics[width=\textwidth,height=7.5cm]{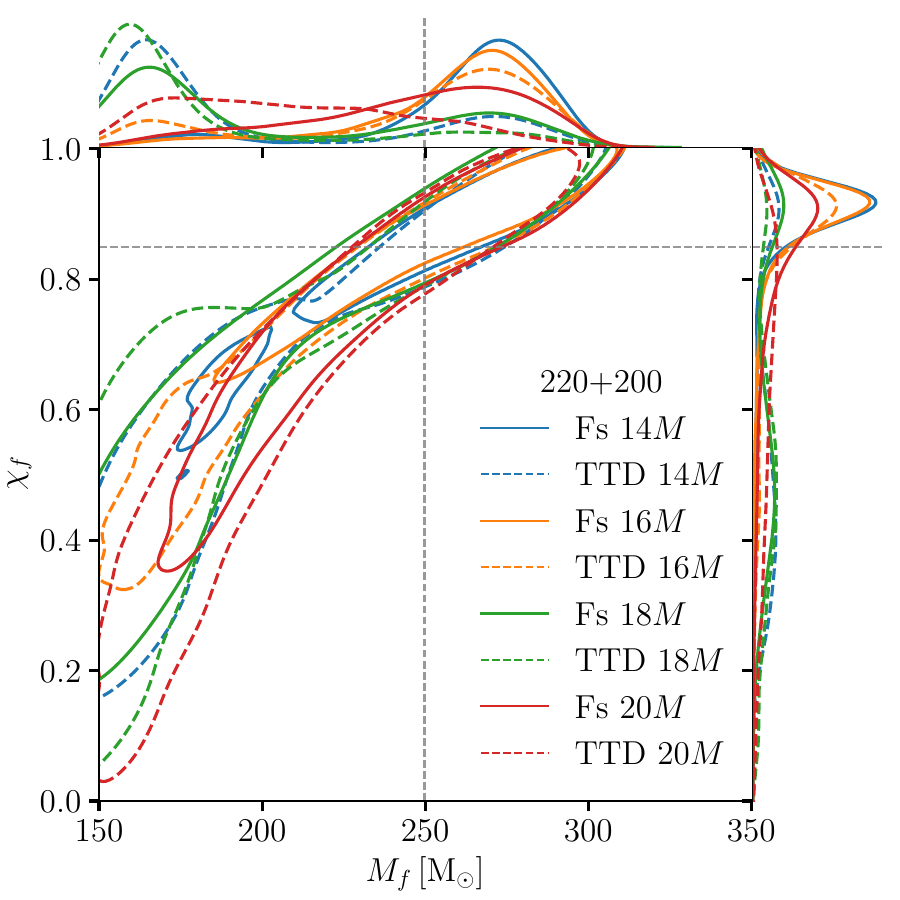}
\end{subfigure}\\
\caption{
Posterior distributions of the final mass $M_f$ and final spin $\chi_f$ recovered from the ringdown analysis of the \ac{NR} injection SXS:BBH:1282. The results are obtained utilizing the $\fs$ method (solid contours) and the \ac{TTD} method (dashed contours). The true remnant parameters of the simulation are indicated by dashed lines. The top panels show results based on the fundamental-mode-only ($220$) hypothesis across start times $\Delta t=8\,M$ to $20\,M$. The middle and bottom panels display the results for the $220+221$ and $220+200$ combinations, respectively. All contours represent the $90\%$ credible level, and the marginal posterior distributions for $M_f$ and $\chi_f$ are shown in the top and right sub-panels.
}\label{fig:fmfs1_sxs}
\end{figure*}

To accurately represent the signal complexity, the injection includes higher-order multipole modes: $(\ell, |m|) = (2,2), (2,1), (2,0), (3,2), (3,3), (4,3), \text{and } (4,4)$. We scaled the waveform to a redshifted chirp mass of $\mathcal{M}_c(1+z) = 108\,M_\odot$, a luminosity distance of $D_L = 3000\,\mathrm{Mpc}$, an inclination $\iota=0.36$ and a phase $\phi=0$. 
The sky location and polarization parameters were fixed to values consistent with the GW231028 ringdown analysis: right ascension $\alpha = 0.04$, declination $\delta = -0.10$, and polarization angle $\psi = 1.27$.

With this configuration, the \ac{SNR} of the full injected \ac{IMR} signal is approximately $23$. The merger produces a remnant black hole with a redshifted final mass of $M_f \approx 249.8\,M_\odot$ and a dimensionless final spin of $\chi_f \approx 0.85$. These remnant properties are in excellent agreement with the values inferred for GW231028, making this an ideal testbed for characterizing the ringdown signature.
As shown in Fig.~\ref{fig:amps1282}, the $(2,1)$ mode of this injection emerges as a significant feature, reaching an amplitude of approximately one-third that of the $(2,2)$ mode at the waveform peak ($t=0$). This prominent relative amplitude confirms that a detectable contribution from the $(2,1)$ harmonic is physically expected for a remnant produced by such a progenitor system.

\begin{figure*}
\centering
\begin{subfigure}[b]{0.48\linewidth}
\centering
\includegraphics[width=\textwidth,height=8cm]{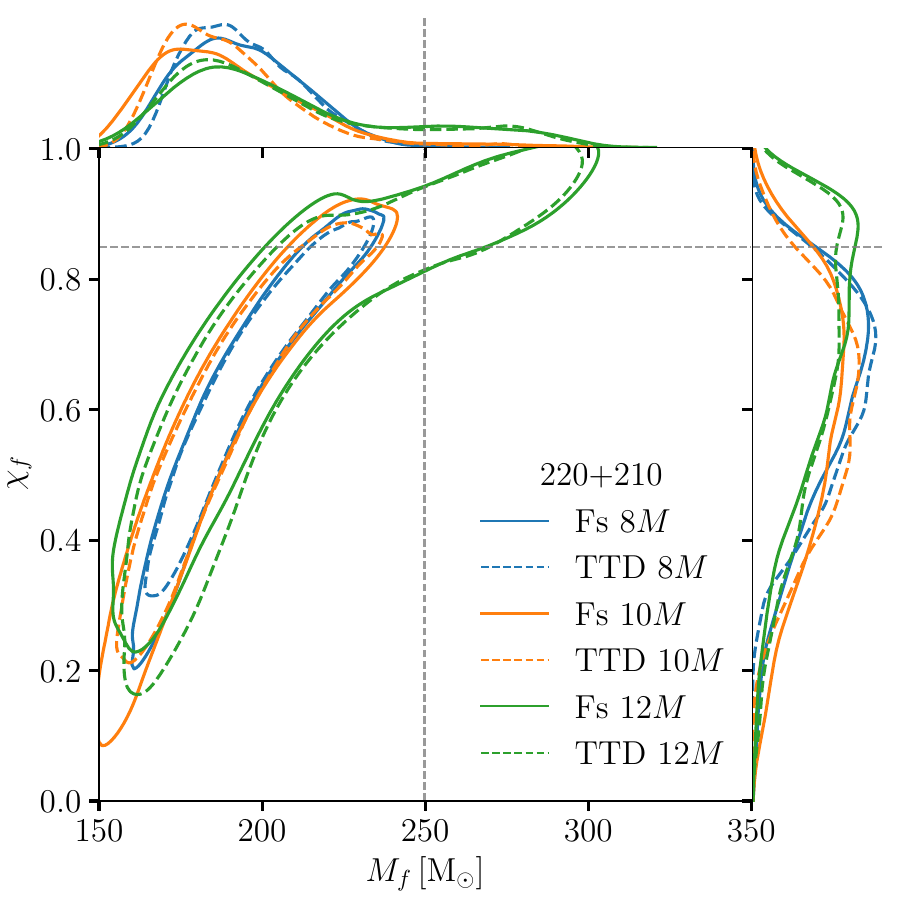}
\end{subfigure}%
\begin{subfigure}[b]{0.48\linewidth}
\centering
\includegraphics[width=\textwidth,height=8cm]{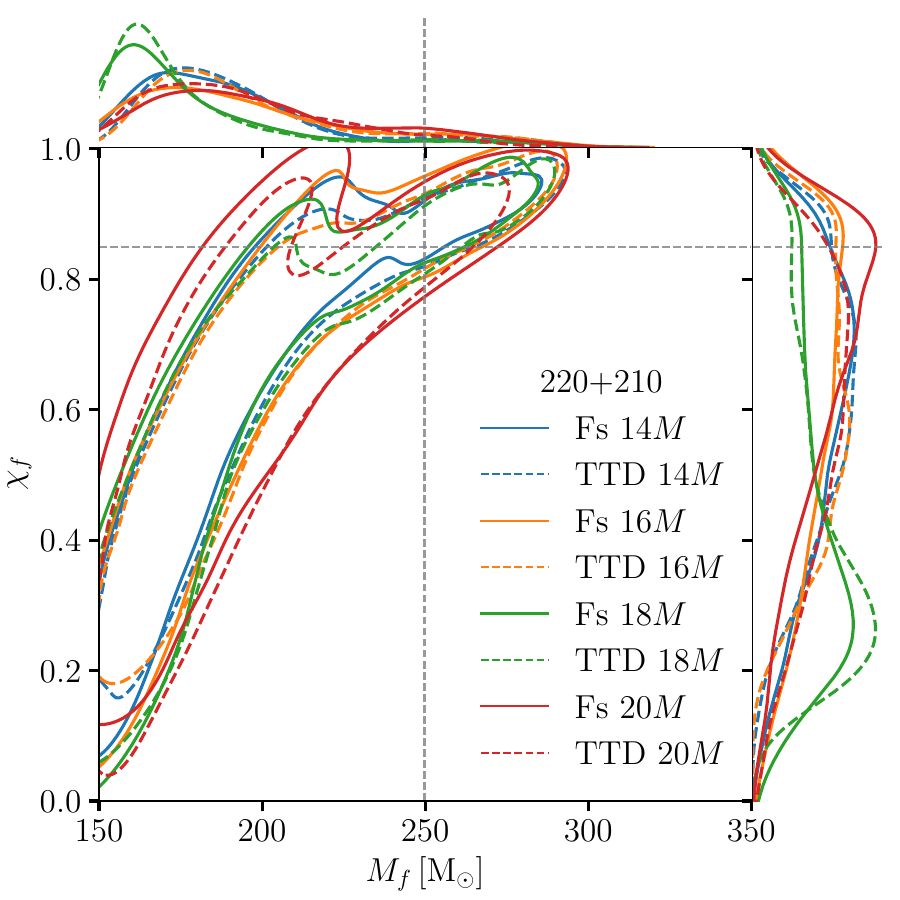}
\end{subfigure}\\
\begin{subfigure}[b]{0.48\linewidth}
\centering
\includegraphics[width=\textwidth,height=8cm]{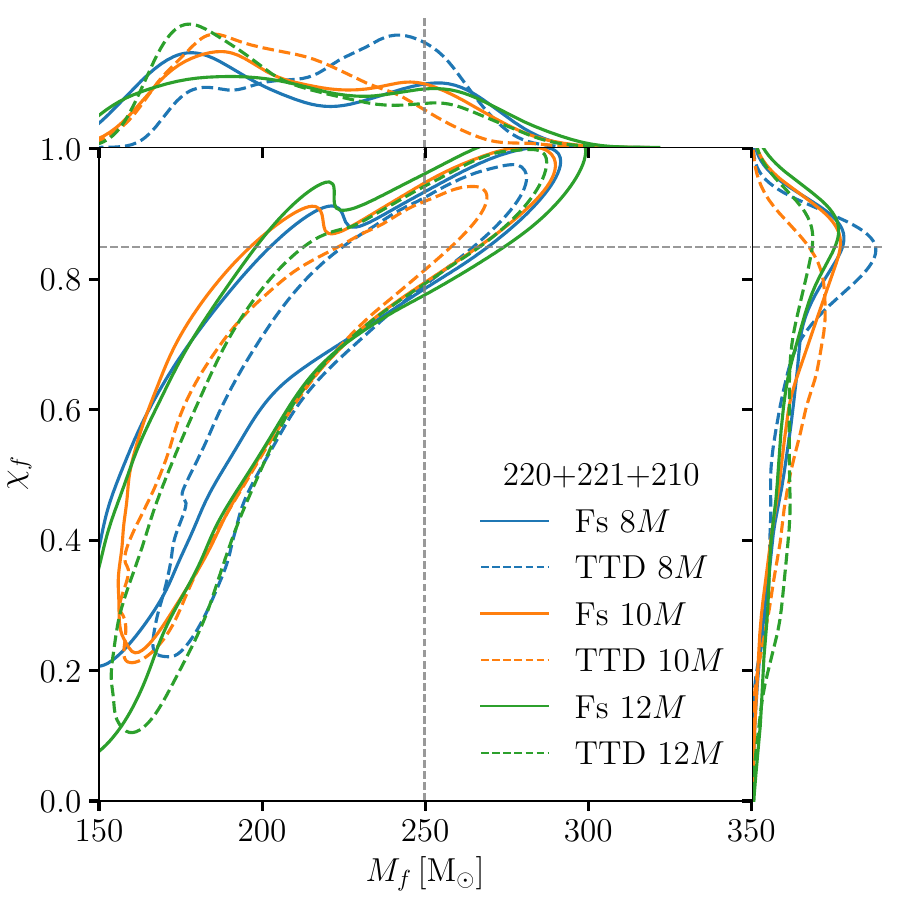}
\end{subfigure}%
\begin{subfigure}[b]{0.48\linewidth}
\centering
\includegraphics[width=\textwidth,height=8cm]{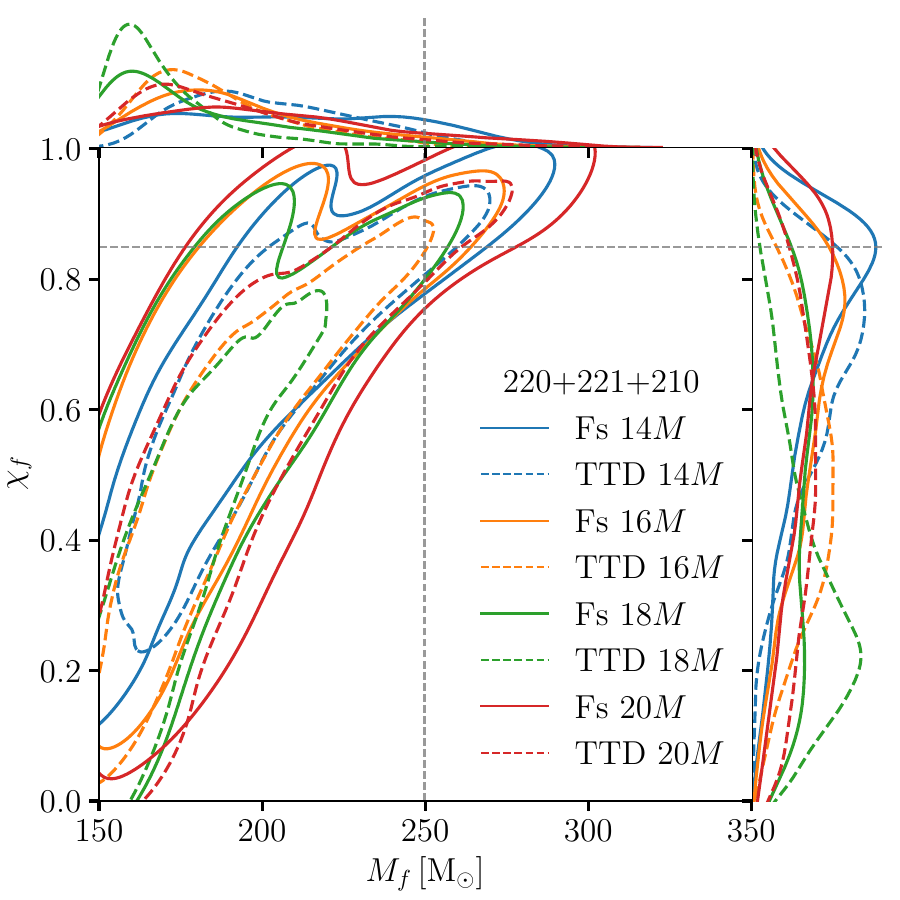}
\end{subfigure}\\
\caption{
Similar to Fig.~\ref{fig:fmfs1_sxs}. 
Posterior distributions of the final mass $M_f$ and final spin $\chi_f$ for the SXS:BBH:1282 injection, inferred from the $220+210$ and $220+221+210$ mode combinations.
}\label{fig:fmfs2_sxs}
\end{figure*}

\subsection{Analysis and Results}
We analyzed the injected data using the same methodology applied to the real event, employing both the $\fs$ method and the \ac{TTD} sampler. To contextualize the ringdown analysis, we first illustrate the time-domain evolution of the injection's mode amplitudes in Fig.~\ref{fig:amps1282}. The results of the Bayesian model selection are presented in Fig.~\ref{fig:logB1282}, and the evolution of the recovered remnant mass and spin posteriors is displayed in Figs.~\ref{fig:fmfs1_sxs} and \ref{fig:fmfs2_sxs}. The injection analysis yielded the following key insights:

\begin{figure*}
\centering
\includegraphics[width=0.88\textwidth,height=15cm]{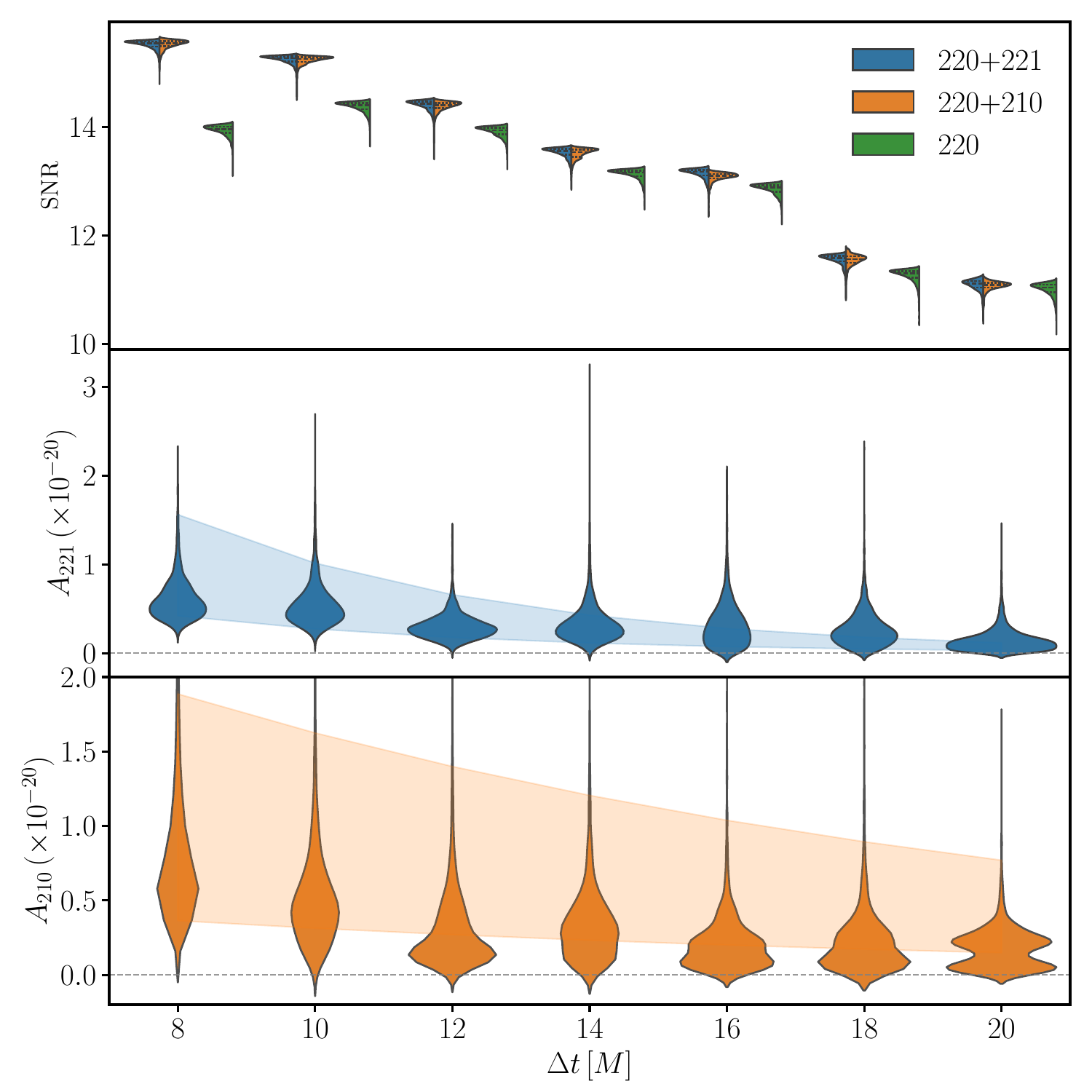}
\caption{
Similar to Fig.~3 in the main text, the top panel shows the posterior distributions of the optimal \ac{SNR} for the $220$, $220+221$, and $220+210$ mode combinations.
The middle and bottom panels display the posterior distributions for the amplitudes of the first overtone ($A_{221}$) and the $(2,1)$ mode ($A_{210}$), respectively, as a function of the start time $\Delta t$.
The theoretical decay bands ($90\%$ credible regions) are anchored to the reference times $\Delta t=10M$ (for $221$) and $\Delta t=8M$ (for $210$).
The striking overlap between the event and the injection demonstrates that the signal morphology of GW231028 is fully consistent with the prediction from a numerical relativity simulation with similar source properties.
}\label{fig:ampSNR1282}
\end{figure*}

\begin{itemize}
\item \textbf{Confirmation of the (2,1) Mode:} The analysis correctly identified the presence of the $(2,1)$ mode. As shown in Fig.~\ref{fig:amps1282}, this mode is a significant sub-dominant feature in the simulation, with an amplitude approximately $1/3$ of the dominant $(2,2)$ mode at the peak time. The recovered Bayes factors (see Fig.~\ref{fig:logB1282}) and parameter estimates (see Fig.~\ref{fig:fmfs2_sxs}) for the $220+210$ model closely track the evolution observed in the real GW231028 data, confirming that the decisive evidence found in the main analysis is consistent with a physical signal of this morphology.

\item \textbf{Consistency of the Overtone:} The pipeline also successfully recovered the first overtone ($221$ mode). The simultaneous high statistical support for the $220+221$ combination in the injection (cf. Fig.~\ref{fig:logB1282}) mirrors the behavior seen in the real event. Furthermore, the $220+221$ model yields unbiased parameter estimates at early times (see Fig.~\ref{fig:fmfs1_sxs}), supporting the conclusion that both physical features—the overtone and the higher harmonic—contribute to the observed signal.

\item \textbf{Sampler Performance on the (2,0) Mode:} The injection study also shed light on the discrepancy observed between the $\fs$ and \ac{TTD} methods regarding the $220+200$ mode combination. In the analysis of SXS:BBH:1282, as illustrated in Figs.~\ref{fig:logB1282} and \ref{fig:fmfs1_sxs}, we found that the standard \ac{TTD} sampling for the $(2,0)$ mode tends to become trapped in local optima due to the frequency proximity of the $(2,0)$ and $(2,1)$ modes. In contrast, the $\fs$ method, by analytically maximizing over amplitudes and phases, is more effective at resolving closely spaced spectral peaks. This validation justifies our preference for the $\fs$ results in cases where the two methods diverge for specific mode combinations.
\end{itemize}

To further validate the physical plausibility of the recovered signal amplitudes, we show the results from the SXS:BBH:1282 injection. 
As evident in Fig.~\ref{fig:ampSNR1282}, the injection analysis reproduces not only the statistical significance (\ac{SNR}) but also the amplitude magnitude and temporal evolution observed in the real event. The recovered amplitudes for both the overtone ($A_{221}$) and the higher harmonic ($A_{210}$) in the injection are quantitatively consistent with those of GW231028. Given that SXS:BBH:1282 represents a fully general-relativistic prediction for a system with similar mass ratio and spin properties, this agreement provides robust evidence that the amplitudes measured in GW231028 are physically expected and not artifacts of noise or sampler systematics.

\nocite{apsrev42Control}
\bibliographystyle{apsrev4-2}
\bibliography{prr}

\end{document}